\documentclass[10pt,journal,compsoc]{IEEEtran}
\usepackage{lipsum}

\newcommand\blfootnote[1]{%
  \begingroup
  \renewcommand\thefootnote{}\footnote{#1}%
  \addtocounter{footnote}{-1}%
  \endgroup
}

\ifCLASSOPTIONcompsoc
  \usepackage[nocompress]{cite}
\else
  \usepackage{cite}
\fi
\ifCLASSINFOpdf
\else
\fi
\usepackage[utf8]{inputenc}
\usepackage[T1]{fontenc}

\usepackage{graphicx}
\graphicspath{ {assets/img/} }

\usepackage{url}
\usepackage[breaklinks]{hyperref}
\usepackage{breakurl}

\usepackage{tikz}
\def\checkmark{\tikz\fill[scale=0.4](0,.35) -- (.25,0) -- (1,.7) -- (.25,.15) -- cycle;} 

\usepackage{subcaption}

\usepackage{algorithm2e}
\usepackage{algpseudocode}
\usepackage{listings}
\usepackage{color}
\definecolor{codegreen}{rgb}{0,0.6,0}
\definecolor{codegray}{rgb}{0.5,0.5,0.5}
\definecolor{codepurple}{rgb}{0.58,0,0.82}
\definecolor{backcolour}{rgb}{0.95,0.95,0.92}
\definecolor{lightgray}{rgb}{.9,.9,.9}
\definecolor{darkgray}{rgb}{.4,.4,.4}
\definecolor{purple}{rgb}{0.65, 0.12, 0.82}

\lstdefinestyle{abgraph_style}{
    backgroundcolor=\color{backcolour},   
    commentstyle=\color{codegreen},
    keywordstyle=\color{magenta},
    numberstyle=\tiny\color{codegray},
    stringstyle=\color{codepurple},
    basicstyle=\footnotesize,
    breakatwhitespace=false,         
    breaklines=true,                 
    captionpos=b,                    
    keepspaces=true,                 
    numbers=left,                    
    numbersep=5pt,                  
    showspaces=false,                
    showstringspaces=false,
    showtabs=false,                  
    tabsize=2
}

\lstdefinelanguage{JavaScript}{
  keywords={typeof, new, true, false, catch, function, return, null, catch, switch, var, if, in, while, do, else, case, break, Trellis},
  keywordstyle=\color{blue}\bfseries,
  ndkeywords={class, export, boolean, throw, implements, import, this, PREFERENTIAL_ATTACHMENT, HEXAGONAL, HYPERCUBE, TRIANGULAR, MESH, objTrellis},
  ndkeywordstyle=\color{codegreen}\bfseries,
  identifierstyle=\color{black},
  sensitive=false,
  comment=[l]{//},
  morecomment=[s]{/*}{*/}{\#},
  commentstyle=\color{darkgray}\ttfamily,
  stringstyle=\color{red}\ttfamily,
  morestring=[b]',
  morestring=[b]"
}
\begin{document}
\title{Bare Demo of IEEEtran.cls for\\ IEEE Computer Society Journals}

\title{\LARGE \bf
AGAPECert: An Auditable, Generalized, Automated, Privacy-Enabling Certification Framework \\with Oblivious Smart Contracts
}

\author{Servio Palacios$^{1}$,~\IEEEmembership{Student Member,~IEEE}, Aaron Ault$^{2}$, James V. Krogmeier$^{2}$,~\IEEEmembership{Member,~IEEE},
\\Bharat Bhargava$^{1}$,~\IEEEmembership{Fellow,~IEEE},
and Christopher G. Brinton$^{2}$,~\IEEEmembership{Senior~Member,~IEEE}% <-this % stops a space
\thanks{$^{1}$Computer Science Department, Purdue University, 305 N University St, West Lafayette, IN, USA
        {\tt\small \{spalacio,bbshail\}@purdue.edu}}%
\thanks{$^{2}$Electrical and Computer Engineering, Purdue University, 465 Northwestern Ave, West Lafayette, IN, USA
        {\tt\small \{ault,jvk,cgb\}@purdue.edu}}%
%\thanks{$^{3}$Agricultural and Biological Engineering, Purdue University, 225 S. University St, West Lafayette, IN, USA
%       {\tt\small dbuckmas@purdue.edu}}%
}

\IEEEpubidadjcol

\markboth{Journal of Dependable and Secure Computing,~Vol.~X, No.~Y, July~2022}%
{Palacios \MakeLowercase{\textit{et al.}}: AGAPECert}



\IEEEtitleabstractindextext{%
%-------------------------------------------------------------------------------
\begin{abstract}
%-------------------------------------------------------------------------------

This paper introduces AGAPECert, an Auditable, Generalized, Automated, Privacy-Enabling, Certification framework capable of performing auditable computation on private data and reporting real-time aggregate certification status without disclosing underlying private data. AGAPECert utilizes a novel mix of trusted execution environments, blockchain technologies, and a real-time graph-based API standard to provide automated, oblivious, and auditable certification. Our technique allows a privacy-conscious data owner to run pre-approved \textit{Oblivious Smart Contract} code in their own environment on their own private data to produce Private Automated Certifications.  These certifications are verifiable, purely functional transformations of the available data, enabling a third party to trust that the private data must have the necessary properties to produce the resulting certification.
Recently, a multitude of solutions for certification and traceability in supply chains have been proposed. These often suffer from significant privacy issues because they tend to take a "shared, replicated database" approach: every node in the network has access to a copy of all relevant data and contract code to guarantee the integrity and reach consensus, even in the presence of malicious nodes. In these contexts of certifications that require global coordination, AGAPECert can include a blockchain to guarantee ordering of events, while keeping a core privacy model where private data is not shared outside of the data owner's own platform. 
AGAPECert contributes an open-source certification framework that can be adopted in any regulated environment to keep sensitive data private while enabling a trusted automated workflow.

\end{abstract}

\begin{IEEEkeywords}
    Oblivious Smart Contract, Private Automated Certification, Certification, Supply Chain, Blockchain.
\end{IEEEkeywords}

}

\maketitle

\IEEEdisplaynontitleabstractindextext

% \fi
\IEEEpeerreviewmaketitle

%-------------------------------------------------------------------------------
\IEEEraisesectionheading{\section{Introduction}
\label{sec:introduction}
}
%-------------------------------------------------------------------------------
\IEEEPARstart{R}{ecently}, many solutions for certifiability and traceability in supply chains using blockchain technologies have risen to prominence~\blfootnote{\copyright 2022 IEEE.  Personal use of this material is permitted.  Permission from IEEE must be obtained for all other uses, in any current or future media, including reprinting/republishing this material for advertising or promotional purposes, creating new collective works, for resale or redistribution to servers or lists, or reuse of any copyrighted component of this work in other works.}.  For example, to provide accountability and visibility in the food supply, blockchain solutions exist such as IBM Food Trust~\cite{IBMFoodTrust}, BeefChain~\cite{Pirus2019}, Lowry Solution’s Sonaria platform~\cite{Lowry2019}, ripe.io~\cite{Ripe2019}, OriginTrail~\cite{OriginTrail2019}, and SAP Blockchain Service~\cite{SAP2019}, to name a few.

A component of many blockchain solutions is a smart contract: a piece of code stored in the blockchain itself that is run by all (or most of) the nodes in the chain in response to some \textit{on-chain} event: some data is added to the blockchain computing nodes which triggers processing. Nodes in the network have copies of the data and the code and achieve data integrity by checking each other's work.  This approach suffers from significant privacy challenges because data and smart contract code is replicated on all nodes in the network~\cite{br2018blockchain} as a means to achieve varying levels of Byzantine fault tolerance, i.e., malicious actors are unable to affect data integrity. 

Thus, some projects have proposed the use of trusted execution environments in the chain to provide some level of privacy-preserving computation~\cite{br2018blockchain} (\S\ref{agapecert_ssec:tees}). A trusted execution environment (TEE) is a private, integrity-protected, and secure computation environment built into processor hardware\footnote{Throughout the rest of this paper, we use the terms TEE, enclave, and \textbf{trusted black box} interchangeably.}~\cite{6799152, Costan2016}. 
For example, Intel SGX (Software Guard Extensions) intends to supply confidentiality and integrity guarantees to computation run in environments where the hypervisor (virtualized environments), operating system, or the kernel are probably malicious/adversarial~\cite{Costan2016}.
TEEs were originally intended to provide a way to keep code and data isolated from malicious actors in a shared platform, and they have had questionable (\S\ref{ssec:intel_sgx_vulnerabilities}) success toward that goal thus far.  However, AGAPECert's trust model simply requires that a TEE can produce cryptographically secure guarantees about which code it ran, not that it isolates data from the rest of its platform.

Brandenburger et al. contributed an open-source proof of concept for TEEs within blockchain computation on top of Hyperledger Fabric~\cite{br2018blockchain}. Another example using blockchain and Intel SGX is SDTE~\cite{8759960}, a data processing model implemented on Ethereum.  A more general framework is the Confidential Consortium Blockchain (COCO) that aims to enable scalable and confidential blockchain networks~\cite{Coco}. In all these solutions, the computation is on-chain, i.e., replicated across nodes in the network.

In many of the real use cases requiring the privacy provided by TEEs, participants are reluctant to provide even encrypted versions of sensitive data to a public or permissioned blockchain which would preserve the encrypted data immutably, thus giving attackers an enormous time-based attack surface to find exploits in encryption key management.

In this paper, we address these challenges by developing \textbf{AGAPECert}, which leverages privacy-preserving computation via TEEs but allows the TEEs to run in environments controlled by the data owner rather than on-chain. AGAPECert abstains from sending data to a public or even private blockchain network and applies restrictions to the code that runs inside enclaves: the code or algorithm that runs on the private data must be pre-approved by both data owner and regulator.

Next, we more formally define the problem that AGAPEcert is designed to solve, and then overview the technology we develop in the rest of this paper.

%-------------------------------------------------------------------------------
\subsection{Problem Definition}
\label{ssec:problem_definition}
\begin{figure}
\centering
\begin{subfigure}[b]{0.40\textwidth}
   \includegraphics[width=1\linewidth]{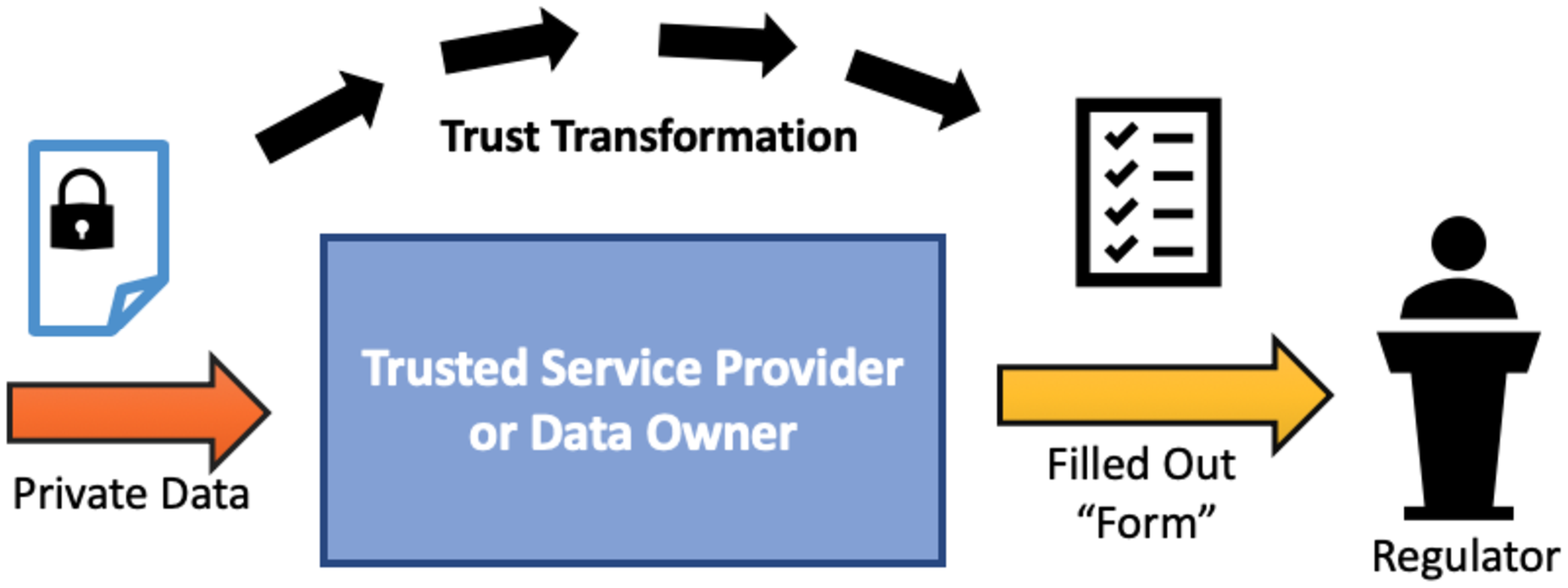}
   \caption{}
   \label{fig:trust_transformation_today} 
\end{subfigure}

\begin{subfigure}[b]{0.40\textwidth}
   \includegraphics[width=1\linewidth]{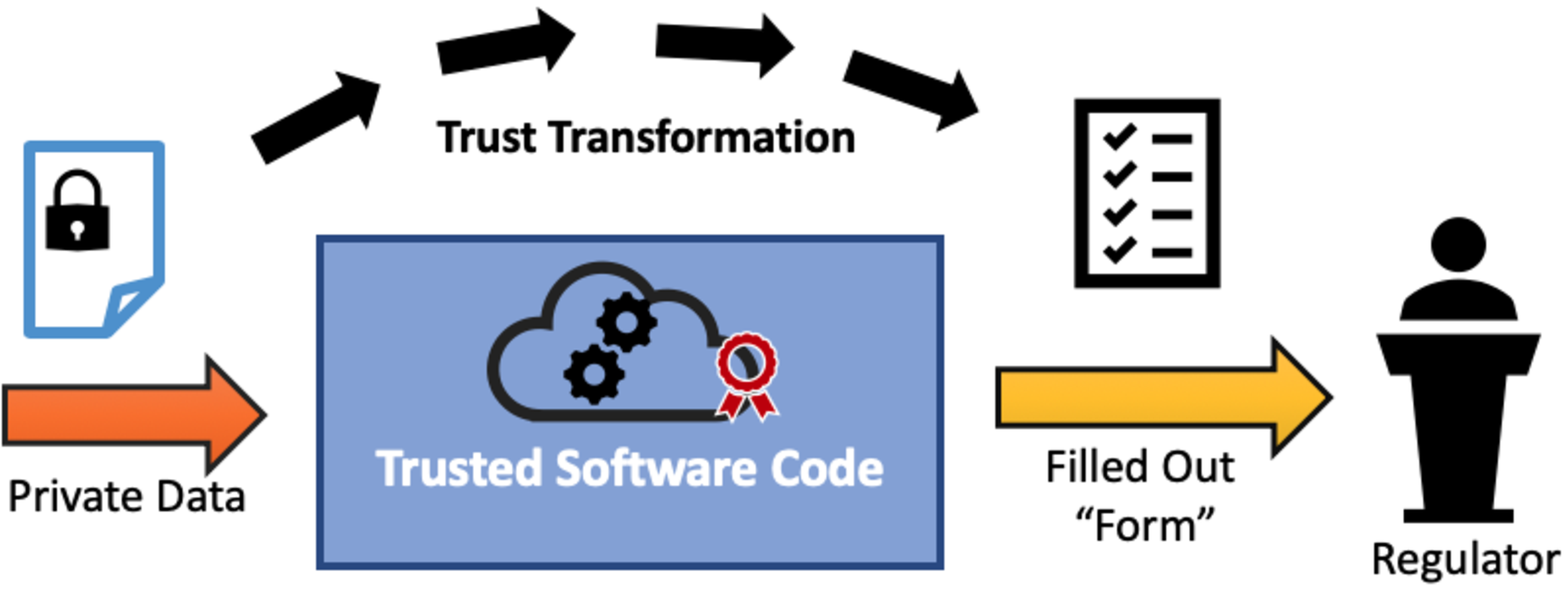}
   \caption{}
   \label{fig:trust_transformation_agapecert}
\end{subfigure}
\caption[Trust Transformation]{(a) Illustration of common model of trust transformation. A person or their trusted agent will fill out a form for a regulator using private, fine-grained source data that they transform into the fields on the regulator's form. (b) AGAPECert replaces or augments the service provider or data owner with a piece of software code agreed upon by both the regulator and the data owner, thus automating the certification process.}
\end{figure}

%-------------------------------------------------------------------------------
% revision 01, 20211010 Figure -> Fig.
Fig.~\ref{fig:trust_transformation_today} illustrates a common user activity that we call a \textit{trust transformation}. A regulator such as the Internal Revenue Service (IRS), an environmental agency, or even a down-stream purchaser of a product has a form that the individual or company being regulated needs to fill out. This form contains a request for information that is derived from other sources. The source data is generally considered private by the data owner, and therefore is not submitted directly to the regulator. The transformation of the source data into the fields on the regulator’s form can be considered a trust transformation from more private, fine-grained data to less private, coarse-grained data.

When the regulator has cause for increased scrutiny, such as an IRS audit or a ``surprise'' inspection, there is typically a need to reproduce the private source data and re-execute the process of the trust transformation under threat of legal action; however, this time under the supervision of the regulator or an independent third party. The source data presented under audit conditions should be verifiable as the same data that produced the responses in the original form. 
The auditor often has little means of verifying that the data provided by the data owner is correct. Instead, the data owner would typically sign some legal document attesting that the data they have provided is correct to the best of his/her knowledge\footnote{The data owner in question is often the sole source of that information.}.

% revision 01 - 20211010 Figure -> Fig.
The goal of AGAPECert is to enable the trusted service provider or data owner that performs the trust transformation in Fig.~\ref{fig:trust_transformation_today} to be replaced or augmented with \textit{code} agreed upon by both the regulator and the data owner as shown in Fig.~\ref{fig:trust_transformation_agapecert}. This automates the process of data-centric certification. This process must not infringe on existing models for trust transformation that society already understands and uses, as outlined below. 

%-------------------------------------------------------------------------------
\subsection{Design Principles}
\label{agapecert_ssec:design_principles}
%-------------------------------------------------------------------------------

The following features must be supported in order for AGAPECert to fit most existing certification processes~\cite{Ray2017,oraclescm,foodsupplychains,chainofcustodycertification,globalgap,supplychaincertificationguide}:

\begin{enumerate}
	\item The data owner should be confident that private data will not be released to the regulator, even in encrypted (but decryptable) form.
 	\item The data owner should be able to run the code to fill out the regulator’s form as often as they like internally without notifying the regulator.
	\item The regulator must be able to verify that the private data has not changed in the event of a subsequent audit under threat of legal action.
	\item The regulator should learn nothing more about the data owner's information or business processes other than the exact features of the filled-out form.
	\item In some cases (\S\ref{agapecert_ssec:trust_levels}, Trust Level 2 and 3), the regulator should be able to confirm that the appropriate code was run without requiring a full audit.
	\item The process need not verify the private data beyond a legal assertion by the data owner that the data is correct.  However, the process may enable better trust requirements around the integrity of private source data.
\end{enumerate}
This functionality leads to repeatable precision: when the trusted software code produces the responses in the form such as ``passed'' or ``properly certified,'' then the data owner knows that they have passed the certification process, regardless of later human regulators or auditors.
The traditional model of data-centric automation has been to ship data to the code that uses it, thus creating privacy concerns. In the AGAPECert model, data-centric automation is achieved by shipping trusted code to the data, eliminating any needless privacy concerns, leaving only those privacy issues required by the contents of the regulator's report (\S\ref{agapecert_sec:example_applications}).

%-------------------------------------------------------------------------------
\subsection{Overview of Technical Approach}
\label{agapecert_ssec:approach}
%-------------------------------------------------------------------------------
As a key component of the AGAPECert framework, we introduce \textit{Oblivious Smart Contracts} (OSCs) as a means to achieve \textit{Private Automatable Certifications} (PACs). A summary of this approach is as follows: 
\begin{enumerate}
	\item Use a piece of standardized, industry-trusted, regulator-approved code that can securely access private data using a standardized graph-based API (\S\ref{agapecert_ssec:trellisfw}).
	\item Compute an aggregate resulting certification (the PAC) as a purely functional result from the input data.
	\item Store results back to the platform of choice for the code (i.e., a blockchain, or any standardized event-ordering scheme~\S\ref{agapecert_ssec:trellisfw}). %graph-based platform). 
	\item Hash and sign all (private data, the result of the computation, and code) so that these signatures and hashes can be presented in the event of a legal challenge, used to verify that the code was run faithfully, and used to verify that the input data has not been changed since code execution. 
\end{enumerate}
No information will be leaked from the private data beyond what is produced by the pre-approved code itself. Even the produced PAC does not necessarily have to be shared by the data owner except in the case of a manual audit or legal challenge. % Review 2021.08.29 -> It need not  -> It needs not
It needs not even leak that the data owner has run the OSC if the OSC itself does not communicate with any outside platform during execution. We can consider the PAC as being produced by a \textit{smart contract} -- standard, pre-approved code shared by participants -- and consumers of the resulting PAC as being \textit{oblivious} to all features of the underlying private dataset beyond the aggregate information in the PAC, as in \textit{oblivious computation}~\cite{Zheng2017}. %These two features give rise to the name we have coined here, \textit{Oblivious Smart Contract}.
As an example of a PAC protocol utilizing blockchain as a byzantine fault tolerant\footnote{A Byzantine Fault Tolerant (BFT) network can continue operating even if some of the nodes fail to communicate or act maliciously.} data storage layer, AGAPECert can leverage auditable computation through a Blockchain-Gateway that allows pluggable shared ledgers (\S\ref{agapecert_ssec:fabric_implementation}) to store anonymous hashes computed during the PAC generation process. 

AGAPECert utilizes the Trellis framework~\cite{Trellis2019}, a graph datastore abstraction which specializes the Open Ag Data Alliance (OADA) API framework~\cite{OADA2019}, to provide a standard API for automated data exchange. AGAPECert uses this concept of having a known, standard API for any type of data as a foundational component to build an interoperable codebase capable of interacting with different individual platforms (\S\ref{agapecert_ssec:trellisfw}.) 
% review: 20210829 Withouta -> Without a
Without a standardizable API layer, it is not practical to write a piece of code that one would expect to work against many heterogeneous data sources. %For those unfamiliar with Trellis and OADA, it is important to note that the API framework works for any kind of data or use case, not simply those within the agriculture industry or the food supply chain.  Therefore, AGAPEcert's use of Trellis and OADA enables it to be fully generalizable.

AGAPECert also integrates two techniques proposed in Intel SGX~\cite{IntelDocumentationSGXIntro}: \textit{REPORT} and \textit{QUOTE}. A \textit{REPORT} is a unique signed structure that binds a key to the enclave hardware, the signer of the codebase, the code itself, and any user-defined data. In the remote attestation process (\S\ref{agapecert_sssec:ra}), the \textit{Quoting Enclave} verifies the \textit{REPORT} and creates and signs the \textit{QUOTE} with a key that is only known to trusted Intel SGX hardware. The \textit{QUOTE} is utilized by the Intel Remote Attestation Services to verify the identity of particular code running inside an enclave. AGAPECert can store the $Quote\_Hash$ (\S\ref{agapecert_ssec:hash_functions}) in the PAC or a shared ledger, serving as BFT proof of the computation result timing.

%2022.02.21
AGAPECert differs considerably from current edge computing literature. The AGAPECert architecture includes a graph data store node, a compute engine node, a broker, and a validator (\S\ref{agapecert_ssec:architecture}). This is a flexible approach: OSC code can interact with or be initiated by regular on-chain smart contract code, various data components can be chosen by participants as either on-chain or off-chain to support the use case, and results and hashes can be reported directly to certifying bodies, pushed to a blockchain (or other standardized event-ordering scheme), or held only by the data owners.   PACs can also be composed: one ``meta''-PAC can be created by an OSC which verifies the validity of many other PACs, avoiding the need to even disclose the underlying PACs themselves.

%-------------------------------------------------------------------------------
\subsection{Trust Levels}
\label{agapecert_ssec:trust_levels}
%-------------------------------------------------------------------------------
\begin{table*}[t]
\centering
\begin{tabular}{|l|l|c|c|c|c|c|}
	\hline
	\textbf{Level} & \textbf{Explanation} & \multicolumn{2}{|c|}{Requirements} & \multicolumn{3}{|c|}{Security Guarantees} \\
	& & 
	\textbf{Intel SGX} & \textbf{Blockchain} & auditable & independently attestated &  provable sequence \\
	\hline
	Trust Level 1 TL1 & Owner Attested      & X         & X          & \checkmark      & X    & X\\
	Trust Level 2 TL2 & Enclave Attested    & \checkmark & X         & \checkmark & \checkmark & X\\
	Trust Level 3 TL3 & Ordering Attested & \checkmark & \checkmark & \checkmark & \checkmark & \checkmark\\
	\hline   
\end{tabular}
\caption{Summary of trust levels in terms of their requirements and security guarantees.}
\label{tab:trust_lelvels}
\end{table*}

Not all use cases have the same trust requirements. AGAPECert proposes classifying OSC structures that solve various use cases into a hierarchy of three trust levels with increasing trust guarantees at the expense of increasing complexity (Table~\ref{tab:trust_lelvels}):
\begin{itemize}
	\item \textbf{Trust Level 1}, Owner Attested (OA): Regulator or consumer of PAC trusts the data owner to faithfully execute the OSC, and therefore does not require proof of correct execution provided by a TEE.  Computation is still auditable under legal challenge. Example: One company prepares a report that utilizes data from a supplier, and they would like to automate report preparation without requiring the transfer of private source data from the supplier that the report preparation process would naturally aggregate anyway.
	\item \textbf{Trust Level 2}, Enclave Attested (EA): Regulator or consumer of PAC requires attestation that OSC code was executed faithfully. Computation is auditable under legal challenge, and correct code execution can be attested without access to private data. Example: A government regulator would like to automate checks against a data owner's digital data, so the resulting PAC contains TEE attestation.
	\item \textbf{Trust Level 3}, Ordering Attested (BA): Regulator or consumer of PAC requires proof of correct execution as well as proof of event ordering for issues like double spending prevention. Contains same components as Level 2, with the addition of a Byzantine fault tolerant shared data storage layer like blockchain.  Example: A buyer of a product wants to know that it is from a certified set (i.e., purchase does not exceed available certified balance), but the seller does not wish to leak information about timing and quantity of other sales.
\end{itemize}

% { revision %2
% 2022.02.14
%AGAPECert exposes these three introductory levels as a backbone that can be extended in the future. Forthcoming articles and software releases of AGAPECert can extend the trust levels according to future events or necessities.
% } revision %2
%-------------------------------------------------------------------------------
\subsection{Contributions}
\label{ssec:contributions}
%-------------------------------------------------------------------------------

The contributions of this paper are summarized as follows:
\begin{itemize}
    \item We develop AGAPECert, an auditable, generalized, automated, and privacy-enabling certification framework that integrates event-ordering technologies, trusted execution environments, and graph-based APIs.
    
    \item As a component of AGAPECert, we introduce Oblivious Smart Contracts and Private Automated Certifications to automate and protect data ownership in real use cases, i.e., certification frameworks, in the supply chain (\S\ref{agapecert_sec:method}).

    \item Through a handshake protocol utilizing trusted execution environments and OAuth2.0 (\S\ref{sssec:PAC_workflow}), AGAPECert contributes auditable computation for use cases that require preserving data ownership and privacy.

    \item We analyze the implications of using trusted execution environments (\S\ref{agapecert_sec:security_analysis}). We also contribute possible extensions that can enhance AGAPECert's future releases. 
    %\item We show that AGAPECert can be used as a novel privacy-preserving food-safety framework.
    \item We provide an open source implementation of our solution (\S\ref{agapecert_sec:prototype_implementation}) that can be reused by other researchers. With this implementation, we demonstrate the key characteristics of AGAPECert in the domain of privacy-preserving food-safety (\S\ref{agapecert_sec:example_applications}, \S\ref{agapecert_sec:evaluation}).
\end{itemize}

%-------------------------------------------------------------------------------
\subsection{Roadmap}
\label{ssec:roadmap}
%-------------------------------------------------------------------------------
The rest of this paper is organized as follows. Section~\ref{agapecert_sec:Background} provides technical background on the components used to build AGAPECert. Then, Section~\ref{agapecert_sec:method} describes the architecture and method through which the AGAPECert system model achieves the goal of private automated certification. Section~\ref{agapecert_sec:security_analysis} provides a security analysis of AGAPECert as well as a discussion on its relationship to known vulnerabilities in Intel SGX. Section~\ref{agapecert_sec:prototype_implementation} describes AGAPECert's implementation details, and Section~\ref{agapecert_sec:example_applications} describes real-world applications that can be deployed using AGAPECert's model. Section~\ref{agapecert_sec:evaluation} presents an evaluation of the most critical of AGAPECert's components, such as trust levels and deployment of OSCs. In Section~\ref{agapecert_sec:related_work}, we compare AGAPECert against the state-of-the-art. Finally, Section~\ref{agapecert_sec:conclusion} concludes this paper. 

%-------------------------------------------------------------------------------
\section{Technical Background}
\label{agapecert_sec:Background}
AGAPECert interacts with a trusted real-time graph-based API (\S\ref{agapecert_ssec:trellisfw}).
AGAPECert computes on encrypted or access-controlled data (confidentiality), preserves the privacy and state of the original private data (integrity, \S\ref{agapecert_ssec:hash_functions}), and, for some use cases, provides the latest record of the certification (sequence, \S\ref{agapecert_sssec:blockchain}). AGAPECert instantiates pre-approved software code inside the compute engine (for Trust Level 2 and 3), providing proof of correct code execution using trusted execution environments and remote attestation (\S\ref{agapecert_ssec:tees}).

%-------------------------------------------------------------------------------
\subsection{Cryptographic Hash Functions and Data Integrity}
\label{agapecert_ssec:hash_functions}
%-------------------------------------------------------------------------------

AGAPECert provides integrity protection of private data and code through well-known cryptographic hash functions. Formally, hash functions map an arbitrary length input message $m$ to a fixed-length output $h(m)$ referred to as a hash~\cite{Menezes:1996:HAC:548089}. The hash has the property that it is computationally infeasible to create an input string which produces a pre-defined hash value, it is infeasible to invert (i.e., determine the original input from the hash alone), and it is deterministic (the same input string always produces the same hash).

AGAPECert uses the SHA256 hashing function~\cite{sha256} to create five hashes (Table~\ref{tab:cryptographic_hashes}) that uniquely characterize the private data: the REPORT (local attestation), QUOTE (remote attestation), PAC, and OSC.

\begin{table*}[t]
\centering
\begin{tabular}{|p{1.8cm}|p{8.0cm}|p{5.3cm}|}
	\hline
	\textbf{Hash Name} & \textbf{Input} & \textbf{Objective}\\
	\hline
	$Data\_Hash$     & $private\_data$ retrieved from a Trellis data store                  & Integrity of Private Data \\
	$Report\_Hash$   & $REPORT$ produced by an enclave when running the OSC                 & Integrity of the REPORT from the enclave\\
	$Quote\_Hash$    & $QUOTE$ produced by a Quoting Enclave                                & Integrity of the QUOTE from the enclave\\
	$PAC_i\_Hash$    & $PAC_i$ (JSON object) produced by the enclave interior               & Integrity of the PAC\\
	$OSC\_Hash$      & $OSC$ Software Code in the Trusted Code Repository                   & Integrity of the OSC code itself\\
	\hline   
\end{tabular}
\caption{Summary of the main AGAPECert cryptographic hashes.}
\label{tab:cryptographic_hashes}
\end{table*}

%-------------------------------------------------------------------------------
\subsection{Trusted Execution Environments}
\label{agapecert_ssec:tees}
%-------------------------------------------------------------------------------

Trusted Execution Environments (TEEs) are an industry innovation to enhance the privacy of data and computation~\cite{br2018blockchain}. Through specialized and isolated  execution environments (enclaves), TEEs shield applications against any malicious operating system, hypervisor, firmware, or drivers~\cite{Costan2016}. TEEs include functionality to encrypt sensitive communications, seal (encrypt) data, and verify the integrity of code and data.  TEEs implementation includes specialized hardware instructions embedded in a machine's processor. Examples of TEEs include Intel SGX~\cite{IntelDocumentationSGXIntro} and ARM TrustZone~\cite{trustzone}.

%-------------------------------------------------------------------------------
\subsubsection{Intel SGX}
%-------------------------------------------------------------------------------
Intel SGX (Software Guard Extensions) aims to supply integrity and confidentiality guarantees through a TEE~\cite{Costan2016}.  Intel SGX creates a private and trusted execution region in the computer's processor called an \textit{enclave}: a secure "virtual container" or black box that contains code and secret data~\cite{Costan2016}. The intended code and data are injected from an \textit{untrusted} region into the enclave. Then, built-in software attestation and sealing mechanisms can provide proof that an application is interacting with the exact/correct software in the enclave and not an attacker's injected malicious code or simulator. 

AGAPECert Trust Level 2 and above require an enclave to exist in the compute engine (\S\ref{agapecert_ssec:architecture}).  The data owner trusts the environment where they run the code on top of their private data, and the code they choose to run there has been pre-approved by them or their trusted service provider in advance.  In addition, the regulator has also pre-approved the code and knows the appropriate \textit{REPORT} parameters that the code will produce when executed in a TEE.  Hence, the code injected in the enclave has been approved by both the \textit{regulator} and the \textit{data owner}, which constitutes a \textit{code trust relationship}. 

%-------------------------------------------------------------------------------
\subsubsection{Local and remote attestation}
\label{agapecert_sssec:ra}
%-------------------------------------------------------------------------------
To prove that specific software code is running in trusted hardware, Intel SGX relies on local and remote attestation~\cite{Costan2016,IntelSDK2019}.  The attestation mechanism provides \textit{proofs}, which comprise a cryptographic signature of the enclave's content (code, data, and parameters) using the platform's secret attestation key known only to the processor. In \textit{local} attestation, the cryptographic proof is verifiable locally by another enclave running in the same processor; this allows secure collaboration to reach a result.

In remote attestation, the cryptographic signature on the proof can be verified by a third party as having originated from a particular piece of trusted hardware using the assumption that the secret key within the processor hardware is unknown outside of the hardware itself and the hardware never reports that key to running software.  In other words, the only entity that could have produced the signature is a trusted processor because it is the only entity that knows its signing key.

AGAPECert utilizes a remote attestation mechanism as the means by which the data owner can prove to the regulator that they have faithfully executed the pre-approved code. 
OSC code reads private data and produces purely functional outputs from that data, including additional hashes (\S\ref{agapecert_ssec:hash_functions}).  This makes code execution both reproducible and verifiable given the same input data.  

%-------------------------------------------------------------------------------
\subsubsection{Intel Enhanced Privacy ID and SGX DCAP}
\label{agapecert_sssec:attestation_service}
%-------------------------------------------------------------------------------

A critical aspect of privacy-preserving computation is attesting that compute devices have not been tampered with and are authentic. Intel's \textit{Enhanced Privacy ID (EPID)}~\cite{IntelEPID2017} is an implementation of ISO/IEM 2008 that handles membership revocation and anonymity.  Membership revocation exposes methods to invalidate compromised secret keys. Anonymity means that EPID will attest to the authenticity of devices without identifying the particular device, i.e., the signature was created by a key from amid a trusted group of secret keys. However, EPID cannot distinguish which particular key in the group created a given signature. 
AGAPECert utilizes these existing remote attestation signature schemes.

EPID has some limitations, however:
\begin{itemize}
    \item Participants are reticent to outsource trust decisions. 
    \item Some highly-distributed use cases require scalable verification points and need to avoid a single point of failure.
    \item AGAPECert can run computation in controlled environments restricting Internet access at runtime.
\end{itemize}
To overcome this, Intel allows the use of Data Center Attestation Primitives -- Intel SGX DCAP~\cite{intelsgxdcap} -- to build customized third party remote attestation. At this point, only servers with Flexible Launch Control (FLC) enabled Intel Xeon E Processors are supported.

%-------------------------------------------------------------------------------
\subsection{Blockchain and Event Ordering Technologies}
\label{agapecert_sssec:blockchain} %OK
%-------------------------------------------------------------------------------

In AGAPECert, we expose three trust levels (\S\ref{agapecert_ssec:trust_levels}) that define different trust requirements. That is, the regulator (or consumer of a PAC) and data owner define the trust requirements for particular use cases. For Trust Level 3, the data owner and regulator have to agree on a technology that serves as a reliable mechanism for the ordering of events. A reliable system must include Byzantine fault-tolerant, consensus, immutability, and integrity properties~\cite{br2018blockchain}. 

Our solution thus desires (for Trust Level 3) a reliable distributed data storage layer for a reduced schema and unique content. 
%In order to maximize the potential for adoption of our solution, we aim to minimize the complexity required for AGAPECert. This led us to propose 
We consider Blockchain as the mechanism for ordering events, given its APIs and platforms have become popular and well-established over the past decade. We will experimentally demonstrate AGAPECert's performance with Blockchain in Section 7.4, where we contribute evidence of a use case deployed in the well-known Hyperledger shared ledger fabric.

A blockchain is an immutable, decentralized digital ledger~\cite{br2018blockchain,Nakamoto_bitcoin:a}. Multiple computers store ordered transactions, linked together through a series of hashes that represent all the data in the ledger up to a given block. Immutability implies that a record in a set "chained" together by hashes cannot be changed without affecting all the subsequent block hashes.
A blockchain provides byzantine fault-tolerant~\cite{Castro:1999:PBF:296806.296824} independent auditability capabilities typically by placing computational constraints on block content which make it too difficult for a malicious attacker to game since they cannot brute-force guess solutions any faster than non-malicious participants. 

A \textit{smart contract} is defined as code that resides in the blockchain itself. An event can trigger some or all nodes to execute that code. The input and output data for each run of the contract code is also typically stored in the blockchain %in order
to make code execution directly verifiable: every node uses the same inputs, runs the same code, and verifies that they produce the same output. 

The simplest forms of OSC do not require a blockchain. However,
some use cases require provable concepts of time or event ordering.  In such cases, including a blockchain building block can be key to giving an OSC the capability to provide collaborative interaction that respects ordering of events. In such cases, AGAPECert can utilize a blockchain to store hashes when building a PAC. AGAPECert currently implements a Generalizable Blockchain-Gateway with IBM Hyperledger Fabric as a building block, but can be extended to other blockchain frameworks such as Ethereum~\cite{Wood2014ETHEREUMAS}.

Alternative solutions to Blockchain for ordering events include employing graph databases 
%Events can be ordered on top of Distributed Databases (e.g., Cassandra~\cite{LakshamAvinash2010}) to expose low-cost services with excellent scalability. For instance, 
% 2022.04.24 Graph Databases 
(such as Neo4j~\cite{Neo4jWeb2} and ArangoDB~\cite{ArangoDB}), which can expose customized ordering of events utilizing network/graph representation of transactions and events. In fact, AGAPECert will employ the Trellis Framework~(\S\ref{agapecert_ssec:trellisfw}) to interact with the private data store via REST API calls, which is built on top of ArangoDB. Hence, as an alternative to Blockchain, one could build personalized ordering of events or traceability modules utilizing the Direct Acyclic Graph (DAG) model materialized on top of ArangoDB and exposed by Trellis via a REST API. 
The distributed ledger technology alternatives to Blockchain, Tangle~\cite{iota} and Hashgraph~\cite{hashgraph9191430}, are also based on DAG models and could similarly be employed here.

%-------------------------------------------------------------------------------
\subsection{Real-time Graph-based API}
\label{agapecert_ssec:trellisfw}
%-------------------------------------------------------------------------------

We utilize the Trellis Framework~\cite{Trellis2019} which exposes standardized REST API semantics to interact with a user's private data store.
The purpose of Trellis is to enable standardized, automated, permissioned, ad-hoc, point-to-point data connections through the use of a common REST API. It is beyond the scope of this paper to fully recount the details of Trellis\footnote{Refer to https://github.com/trellisfw for more information.}.  However, some critical features of Trellis are important to the development of AGAPECert:
\begin{itemize}
    \item \textit{Resource discovery}: Filesystem-like graph schemas define where data can be discovered. For example, catch locations for a fishing vessel for May 1, 2020 could be defined as discoverable at graph path /bookmarks/trellis/fishing/catch-locations/day-index/2020-05-01.
    \item \textit{Write semantics}: Trellis standardizes how data within a graph is written. All data changes are reduced to an ordered stream of idempotent merge operations\footnote{An idempotent merge operation means that a given JSON document is produced that only affects matching keys.  Keys that do not exist are created, existing ones are deep-replaced at overlapping key paths, similar to a common upsert.  Applying the same merge repeatedly results in the same resource state at the mentioned key paths.}. Operation ordering is only guaranteed per resource, not globally.
    \item \textit{Change feeds}: Clients can register for real-time change feeds for any arbitrary subgraph of data.  This provides both a real-time communication channel as well as a means of concurrency-safe 2-way data synchronization.  The change feed is comprised of the ordered stream of idempotent merge operations.
    \item \textit{Authorization}: Trellis standardizes how any client registers and obtains authorization tokens at any Trellis platform.
    \item \textit{Permissions}: Trellis standardizes how data can be locally shared within a platform.
\end{itemize}

%-------------------------------------------------------------------------------
\section{Method: AGAPECert System Model}
\label{agapecert_sec:method}
%-------------------------------------------------------------------------------

%This section describes the System Model and Architecture. 
%We utilize the concepts described in section \S\ref{sec:Background} to motivate the security guarantees of AGAPECert. 
%-------------------------------------------------------------------------------
\subsection{AGAPECert Architecture}
\label{agapecert_ssec:architecture}
%-------------------------------------------------------------------------------

AGAPECert considers two primary actors: \textit{data owners} and \textit{regulators}. Data owners are clients that own private and sensitive data. Regulators are actors that desire some derivative of the client's private, sensitive data without requiring the disclosure of that data itself.  Note that the regulator may not be only what is traditionally considered a regulator, e.g., from a government agency, but rather is used in a broader sense here as any entity looking for information that may be derived from a client’s private data. By this definition, a regulator could be a direct customer of the data owner, a down-stream buyer in a supply chain, or a business partner.  

We define two critical components of the certification process: 
\begin{itemize}
    \item \textit{Private Automated Certification (PAC)}: The derivative output of the client’s data (i.e., the contents of the "form" that the regulator requires the data owner to fill out).
    \item \textit{Oblivious Smart Contract (OSC)}: The regulator-approved code which, given access to private input data, produces the desired PAC (i.e., the "questions" on the regulator's "form").
\end{itemize}  

The client (or their trusted service provider or industry consortium) ensures that the OSC obtains only the necessary data to produce a PAC, and that the PAC will not leak any unauthorized information (such as copies of the private data, or knowledge of when the OSC code was run). The client runs the approved code and provides it access to their private data to it to obtain a signed PAC. This PAC should contain, at minimum, a cryptographic hash representing the input data used in its computation.  For Trust Levels 2 and above, it must also include the cryptographic hashes of REPORT ($Report\_Hash$) and QUOTE ($Quote\_Hash$) from the Intel SGX enclave. The client then provides their PAC to the regulator upon request, at which time the regulator may validate the PAC according to the trust level for that use case. Should a subsequent legal challenge be necessary, the client can produce the private data to a legal authority, which can verify that the cryptographic hash for that data ($Data\_Hash$) matches that from the PAC, and can also re-run the OSC code to produce an equivalent PAC for comparison.  

\begin{figure}[t]	
	\centering
	\includegraphics[scale=0.70]{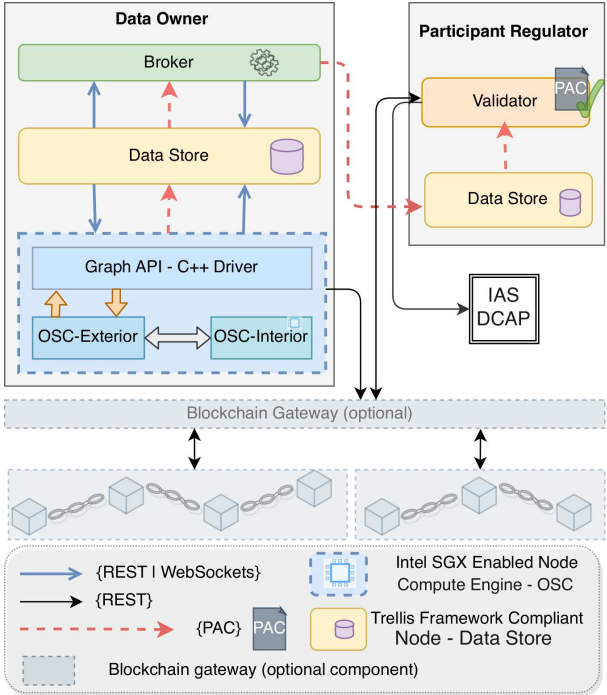}
	\caption{AGAPECert architecture for Trust Level 3. The data owner main components include a compute engine that runs the OSC (exterior and interior), a data store, a service manager for the OSCs (broker), and a blockchain-gateway (optional, for trust level 3 only) to store the $Quote\_Hash$ and UUID (PAC.id). The regulator includes the validator and its data store. The validator queries the ledger to verify a particular PAC. Also, the validator attests correct code execution connecting to remote attestation services or DCAP.} 
	\label{fig:agapecert_architecture}
\end{figure}

% revision 01 - 20211010 Figure -> Fig.
The AGAPECert architecture is comprised of six main components, four required and two optional based on the level of trust required, as depicted in Fig.~\ref{fig:agapecert_architecture}:

% revision 1 - 20211002
\begin{table*}[t]
\centering
\begin{tabular}{|l|l|}
	\hline
	\textbf{Components} & \textbf{Explanation} \\
	\hline
	Compute Engine & Compute node controlled or trusted by the data owner. Runs the OSC. \\
	Data Store     & A Graph Data Store that holds the private data (Trellis).\\
	Broker         & A web-app that initiates, authorizes, provisions, moderates, monitors, validates OSC execution. \\
	Validator      & A web-app that can validate a given PAC \\
	Attestation Service & The Intel SGX attestation service or Data Center Attestation Primitives (DCAP) \\
	Blockchain-Gateway & A Generalizable Blockchain Service to connect to a mix of ledgers as needed (optional). \\
	\hline   
\end{tabular}
\caption{Summary of components comprising the AGAPECert architecture. The Attestation Service and Blockchain-Gateway are only necessary for Trust Level 3.}
\label{tab:components}
\end{table*}

\begin{itemize}
	\item \textbf{Compute Engine}: A compute node controlled by or trusted by the data owner that is capable of running the OSC code. A compute engine is required for all trust levels. For Trust Levels 2 and above, the compute engine must be Intel SGX-enabled.
	\item \textbf{Data Store}: A Trellis-conformant data storage platform that holds the private data owned by the client.  This serves as the source of the data for the OSC, as the real-time communication channel for the broker and the OSC, and as the destination for the PAC produced by the OSC.
	\item \textbf{Broker}: A web application that initiates, authorizes, provisions, moderates, monitors, and validates OSC execution, communicating with the OSC through the secure shared Trellis connection.  This serves as the bridge between the data owner and the OSC, acting as a service manager; all OSC services can be monitored through this web-app.
	\item \textbf{Validator}: A web application that can validate a PAC, including remote attestation for Trust Level 2 and 3, and checking a blockchain (or other event-ordering technology) for Trust Level 3.
	%, as shown in Figure~\ref{fig:broker}. 
	\item \textbf{Attestation Service}: The Intel SGX attestation service or DCAP (\S\ref{agapecert_sssec:attestation_service}). %or Data Center Attestation Primitives (DCAP). 
	Given a particular QUOTE, this service can attest whether the QUOTE was produced by a legitimate Intel SGX enclave or not, thus attesting proper code execution. 
	% 2019.10.17 - Servio - Text for the camera ready
	%Note that this attestation service can be consulted at any time after the QUOTE is produced, not necessarily during execution, thereby enabling the PAC to be stored in the client’s data store until the client decides to disclose the PAC to the regulator. 
	Required for Trust Levels 2 and above. 
	% revision 1 - 20211002
	% revision 2 - 20220302
	\item \textbf{Blockchain (Event-Ordering) Gateway (optional)}: An interface to a blockchain (or other event-ordering scheme, see \S2.3)  that is trusted by the regulator and the data owner---this component is optional and only necessary for trust level 3.
\end{itemize}
These components are summarized in Table \ref{tab:components}.

% revision 01 - 20211002
% AgapeCert Architecture

%-------------------------------------------------------------------------------
\subsection{Oblivious Smart Contracts}
\label{ssec:OSC_implementation}
%-------------------------------------------------------------------------------

An Oblivious Smart Contract (OSC) is software code that reads private data to compute a result such as pass/fail and generate a PAC (\S\ref{sssec:PAC_workflow}). OSCs run in the compute engine node. For Trust Levels 2 and 3, this computation happens inside a TEE on the compute engine. 

As explained by Intel's documentation~\cite{IntelDocumentationSGXIntro}, Intel SGX applications (such as Trust Level 2 and above OSC's) are comprised of two parts: the \textit{untrusted part} of the application which communicates with the enclave, and the \textit{trusted part} that includes the computation inside enclaves. Note that while these terms make sense in the traditional environments where Intel SGX is intended to run, they are misnomers in our context where the environment running the OSC is assumed to be trusted by the data owner already. We will instead use the term \textit{enclave exterior} to describe what Intel SGX terms the untrusted part, and \textit{enclave interior} to describe the trusted part.

Trust Level 1 does not require a TEE; this simple form of OSC is simply any code capable of interacting with a Trellis platform to read data and save a resulting PAC. 

For the Trust Levels 2 and 3, the OSCs include native C Bridge functions that communicate with the enclaves. The enclave exterior of the OSC connects and retrieves the private data from the Trellis data store and injects a buffer into the enclave interior of the OSC. AGAPECert includes C++ classes that implement the Trellis REST and WebSockets APIs to communicate with the Trellis data store through a shared resource located in the Trellis graph at $/bookmarks/OSC/H_k$, where $H_k$ can be a random string generated at runtime by the OSC or a static feature of the OSC and is discovered by the Broker when initiating the connection. 

The enclave interior of the OSC computes the cryptographic hashes necessary to audit the computation in the future and passes them through the enclave exterior to be stored in the PAC, which is stored back to the data owner's Trellis data store. In Trust Level 3, AGAPECert also stores the cryptographic hashes in the blockchain. 

It is important to note that since AGAPECert stores only the cryptographic hashes in the blockchain, and these hashes cannot be linked to the source data using only the hash, this protects the privacy of the data owner; i.e., there is no information leakage from this process such as third party knowledge of how many times the data owner has run the OSC.  However, storage of the hash in the blockchain can leak the time when a given PAC was generated since a regulator receiving the PAC in the future can check where in the blockchain the hash was saved.  

The typical OSC runs continuously, awaiting notification from the broker through the Trellis data store at shared storage location $/bookmarks/OSC/H_k$ to start a new PAC generation process on a new subset of data, or until a restart is submitted from the broker.  Each run of the OSC produces one or more PACs deterministically, after which point the OSC returns to a listening state awaiting further signals %data or instruction 
from the broker.

%-------------------------------------------------------------------------------
\subsection{Private Automated Certifications (PAC) Workflow}
\label{sssec:PAC_workflow}
%-------------------------------------------------------------------------------

\begin{figure*}[!tbp]
\begin{subfigure}[b]{0.40\textwidth}
  	\includegraphics[scale=0.40]{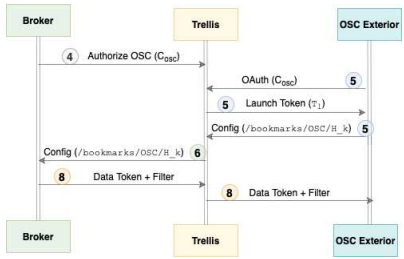}
	\caption{AGAPECert Workflow (Before PAC Generation.)}
	\label{fig:agapecert_workflow_1}
\end{subfigure}
\hfill
\begin{subfigure}[b]{0.50\textwidth}
   \includegraphics[scale=0.40]{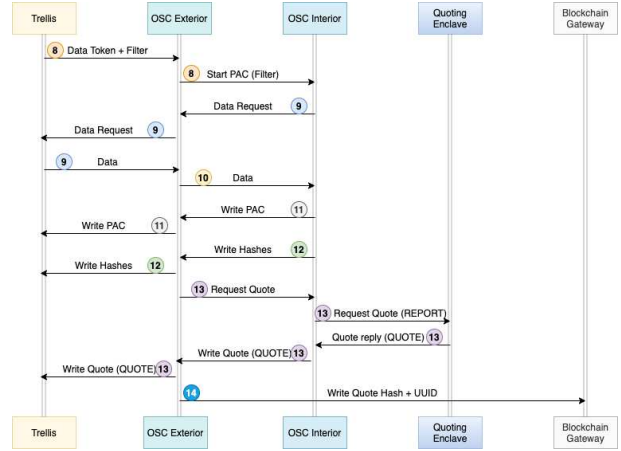}
	\caption{AGAPECert Workflow - PAC Generation Overview.}
	\label{fig:agapecert_workflow_2}
\end{subfigure}
\caption[AgapeCert Workflow]{(a) Illustrates the OSC's Authorization to the provisioning of data (steps 4-8) (b) AGAPECert generates a PAC (steps 8-14).}
\label{fig:agapecert_workflow}
\end{figure*}

To generate a certification under Trust Level 3, AGAPECert uses the following workflow % review 2021.08.29 (Figure~\ref{fig:agapecert_workflow}.) -> (Fig.~\ref{fig:agapecert_workflow}.)
(Fig.~\ref{fig:agapecert_workflow}, Algorithm~\ref{alg:pac_generation_overview}.)  For other trust levels, the respective components not used by those levels are simply left out.  Note that AGAPECert utilizes a set of RFCs in this process (RFC7591, RFC7517, and RFC7519) as prescribed in the Trellis authorization protocol~\cite{trellis_auth}.

\begin{enumerate}
	\item \textit{Install OSC}: The data owner installs the OSC on their compute engine.  This produces a random public/private asymmetric key pair, with the public key saved  as a JSON Web Key (jwk from RFC7517) in a newly generated Trellis Client Certificate ($C_{osc}$).

    \item \textit{Authorize Broker}: The data owner logs into their Trellis compliant node via OAuth2 to authorize a token for the Broker. 
	    
	\item \textit{Watch for OSCs}: The broker opens and maintains an active websocket connection to the Trellis data store that watches the top-level $/bookmarks/OSC$ document for any connected OSCs. 
	
	\item \textit{Authorize OSC}: The data owner uses the Broker to pre-register the OSC's Trellis Client Certificate $C_{osc}$ at their Trellis data store as an authorized OSC.
	    
	\item \textit{Start the OSC}: The data owner starts the OSC. They can also verify that the hash of the OSC code $OSC\_Hash$ matches the one available in a private certified code repository.
	    \begin{enumerate}
	        \item The OSC Exterior performs OAuth2 dynamic client registration by exchanging its Trellis Client Certificate $C_{osc}$ with Trellis for a Client ID.  It then performs OAuth2 Client Grant flow during which it proves that it has the private key for the pre-registered certificate by creating a signed $jwt\_bearer$ token (RFC7523).  This process results in a properly scoped launch token ($T_l$) to access the user's Trellis data store at $/bookmarks/OSC$.
	        \item The OSC Exterior generates a random string $H_k$ that uniquely identifies this instance of OSC. The OSC uses the token received from previous step %\ref{itm:provisiontoken}
	        to create a resource ($/bookmarks/OSC/H_k$) in the data store. It also puts information about itself into that document.
	        \item The OSC exterior opens and maintains an active Trellis websocket connection to the Trellis data store watching for changes to the new $/bookmarks/OSC/H_k$ as the main communication channel between the broker and the OSC.
	    \end{enumerate}
	
	\item \textit{Communication Channel Opens}: The broker’s active Trellis websocket connection notifies it that a new OSC resource exists at $/bookmarks/OSC/H_k$.
	
	%revision 2021.09.19 deleted this: Optional:
	%\item \textit{Validate OSC Quote}: In the event that the data owner requires confirmation that their platform has loaded the correct OSC code, they can load their own credentials for the Attestation Service~\S\ref{agapecert_sssec:attestation_service} (IAS or DCAP) into the broker and it will initiate remote attestation to verify that the enclave is legitimate. In most cases the user already trusts their own platform and would not require signing up with IAS. This remote attestation workflow produces a QUOTE (\S\ref{agapecert_sssec:ra}).
	
	%revision 2: 2022.02.12 rephrase the paragraph to make this required
	\item \textit{Validate OSC Quote}: To validate that their platform has loaded the correct OSC code, the data owner loads the credentials for the Attestation Service~\S\ref{agapecert_sssec:attestation_service} (IAS or DCAP) into the Broker. The Broker will initiate remote attestation to verify that the enclave is legitimate. 
	% 2022.03.02 In most cases, the user already trusts their platform. 
	This remote attestation workflow produces and validates a QUOTE (\S\ref{agapecert_sssec:ra})\footnote{If the OSC cannot be validated, an exception report must be stored in the data owner's trusted platform. This constitutes a remote attestation failure, in which case the process will not continue with the next steps.}. The data owner can store the QUOTE to expose auditability features. The data owner can also prove to a third party (e.g., a regulator or auditor) that a particular OSC was run in the data owner's platform with a specific configuration.

	\item \textit{Provision Data to OSC}: The broker then provisions a properly-scoped data access token $T_d$ for the OSC to use in creating its PAC. The broker writes this token to the shared Trellis communication channel at $/bookmarks/OSC/H_k$ along with any data filtering instructions (such as restricting the PAC to only consider a particular day’s dataset).  The active websocket connection held by the OSC exterior notifies it of the newly provisioned token and filter, triggering the OSC to begin the core PAC generation.
	
	\item \textit{OSC Interior Requests Data}: The OSC exterior receives the data access token and filtering instructions and notifies the OSC interior to begin PAC generation. The OSC interior uses its knowledge of the known, published Trellis semantic data structures to begin requesting data it needs. Requests for data initiated by the OSC interior are forwarded to the OSC exterior to make the actual requests over the active Trellis websocket connection.
	
	\item \textit{OSC Exterior Injects Data}: The OSC exterior serializes and injects the received data into the OSC Interior as it comes back from Trellis. 
	
	\item \textit{OSC Interior Computes PAC}: The OSC interior computes its core certification result (i.e., pass/fail) from the received input data, as well as a cryptographic hash ($Data\_Hash$) of all serialized data received during the generation of one PAC. Upon completion of all received data, the OSC interior saves this hash of all input data to the PAC in the data store.
	
	\item \textit{OSC Interior Hashes PAC}: The OSC interior generates a hash of the entire PAC (including $Data\_Hash$ and a random universally unique identifier for the blockchain transaction) and saves this back to the PAC itself in the Trellis data store through the OSC exterior.
	
	\item \textit{Exterior Obtains TEE Quote}: The OSC exterior sees the hash of the overall PAC and concludes that the OSC interior has completed its work.  The OSC exterior then instructs the OSC interior to obtain a QUOTE from a local quoting enclave, including the hash of the PAC as the user data for a new REPORT from the OSC interior. The OSC exterior gathers the final QUOTE from this process and writes it to the PAC in the Trellis data store. 
	
	\item \textit{OSC Interior Sends $Quote\_Hash$ to Blockchain}: For Trust Level 3, the OSC exterior will then communicate with the Blockchain Gateway to record the unique identifier and $Quote\_Hash$ in the blockchain, along with any further case-specific requirements.
	
	\item \textit{Data Owner Sends PAC to Regulator}: Finally, at a later time, the regulator receives the generated PAC from the data owner and uses the Validator to check it.  The Validator sends the QUOTE from the PAC to an attestation service to verify that the QUOTE was indeed generated with an Intel EPID key that has not been revoked.  Note that if a processor’s key is revoked, this either invalidates all prior PAC’s generated by that processor, or some outside means of providing a trusted timestamp of QUOTE generation (such as that provided by Trust Level 3 through a blockchain) must be included in this flow to maintain validity of PAC's generated prior to some known revocation time. The Validator also queries the blockchain ledger to validate the $Quote\_Hash$ and case-specific data.
\end{enumerate}

% AGAPECert Workflow

%    review 2021.09.18  ==============================================================================================================================================
\SetKwComment{Comment}{/* }{ */}
\begin{algorithm}[hbt!]
\caption{AGAPECert PAC generation overview.}\label{alg:pac_generation_overview}
\KwData{$data\_token,\; filter$}
\KwResult{$PAC,\; QUOTE,\; QUOTE\_HASH,\; UUID$}
%\Comment*[r]{instantiates a new trellis object}
$trellis \gets new \; Trellis(data\_token)$ \Comment*[r]{instantiates a new gateway}
$blockchain\_gateway \gets new \; Blockhain()$ \Comment*[r]{retrieves private data}
$data\gets trellis.getPrivateData(filter)$ \Comment*[r]{instantiates the trusted algorithm inside enclave (OSC interior), returns the computed PAC}
$pac \gets computeAlgorithm(data)$ \;
\If{$pac \neq null$}{ 
    $quote \gets getQuote()$ \Comment*[r]{retrieves the cryptographic proof of the computation}
    $trellis.putPAC(pac, quote)$ \Comment*[r]{updates PAC in the trellis backend}
    $blockhain\_gateway.putPAC(pac)$ \Comment*[r]{stores UUID and Quote\_Hash in the shared ledger}
}
\end{algorithm}

%-------------------------------------------------------------------------------
\subsection{Blockchain-Gateway Schema}
\label{ssec:hyperledger_fabric_schema}
%-------------------------------------------------------------------------------
%As discussed in previous sections, 
AGAPECert stores a minimal set of cryptographic hashes (\S\ref{agapecert_ssec:hash_functions}) in the shared ledger. These hashes do not convey any identifiable private information to an eavesdropper of the transactions in the blockchain nodes. We define the methods of the smart-contract necessary to store hashes from the OSC. 

The AGAPECert proof-of-concept models the business network utilizing Hyperledger Fabric (\S\ref{agapecert_ssec:fabric_implementation}.)  Hyperledger Fabric requires the definition of assets, participants, and transactions. AGAPECert utilizes a PAC as an asset in the blockchain that we define as a "Fabric PAC" $fabPAC$. 
We define two participants: an anonymous participant that stores PACs in the ledger and a regulator who queries the transactions and assets in the distributed ledger.

AGAPECert defines a simple schema of at least two strings for a $fabPAC$: (unique identifier, $Quote\_Hash$) and for some use cases a One-Time-Use-Key ($OTK$).
AGAPECert's blockchain schema is shown in Table~\ref{tab:pac_business_network}. This forms the basis for use cases in Trust Level 3.

\begin{table}[ht]
\centering
\begin{tabular}{|l|l|}
	\hline
	\textbf{Property(fabPAC.)} & \textbf{Type}.   \\
	\hline
	$id$                          & String (UUID) \\
	$quoteHash$                   & String  \\
	$OTK$ (optional One-Time-Key) & String (Base64 encoded) \\
	\hline   
\end{tabular}
\caption{PAC business network in the shared ledger.}
\label{tab:pac_business_network}
\end{table}

% review 2021.09.19
The regulator automated software can check if the cryptographic hashes in the PAC match the transaction registered in the blockchain. AGAPECert's Blockchain Gateway exposes a REST API to communicate with the Fabric.

%-------------------------------------------------------------------------------
\section{Security Analysis}
\label{agapecert_sec:security_analysis}
%-------------------------------------------------------------------------------

AGAPECert does not alter the existing real-world model of requiring the regulator to trust the data provided to it by the client under threat of legal recourse, as discussed in \ref{agapecert_ssec:design_principles}.  Our security analysis therefore only focuses on guarantees made about computation on the private data (which is assumed correct until audited), rather than about the private data itself.

%-------------------------------------------------------------------------------
\subsection{Side channel attacks to the Compute Engine}
\label{ssec:intel_sgx_vulnerabilities}
%-------------------------------------------------------------------------------
TEE technology such as Intel SGX can be vulnerable to "side channel attacks:"~\cite{foreshadow,meltdown,ChenGuoxing2019SSIS,Obliviate2018,XiaoYuan2017SDAS,WangWenhao2017LCot} malicious code running on the same processor can learn about enclave computation and data via round-about methods such as tracking cache timings after an enclave is switched out of execution.  AGAPECert assumes the data owner is running the OSC in an environment they already trust: the threat of a malicious entity on the same processor does not apply, hence AGAPECert is immune to traditional enclave side-channel attacks.

However, a data owner using past side-channel attacks against their own trusted enclaves (such as the enclave that provides quotes) could learn the remote attestation keys~\cite{foreshadow,ChenGuoxing2019SSIS} for their own system and use that to forge fake QUOTE's.  Since such attacks have been discovered, researchers have also contributed mitigation techniques to patch those security vulnerabilities~\cite{foreshadow,ChenGuoxing2019SSIS}.  Additionally, Intel's security advisories provide critical mitigation techniques~\cite{intelsecurityadvisory}. For instance, some vulnerabilities require microcode level and software mitigations. 

AGAPECert runs in a controlled environment in which standards and best practices are enforced; therefore, data owners and participants in the protocol must have the latest Intel SGX Software installed, the newest microcode updates through BIOS updates, and any other recommended measures such as updated operating systems and virtual machines. Microcode updates upgrade the Security Version Number (SVN) utilized in the implementation of Intel SGX~\cite{intelsecurityadvisory}. Microcode updates provide new sealing and attestation keys to the enclaves on the platform~\cite{intelsecurityadvisory}. Hence, we can track the SVN embedded in the PAC the regulator will assess the validity of a specific PAC via a customized attestation process using current vulnerabilities.  

This means that at any given time, it is not known to be possible to forge QUOTEs that contain the latest SVN until a future vulnerability is released.  In the event of catastrophic security failure of Intel SGX's architecture, data owners may need to refresh relevant PACs after updating their processor microcode to produce updated SVNs.  The purely functional and auditable nature of AGAPECert's OSCs fit this model well.  In addition, if the QUOTE hash was published in a blockchain prior to vulnerability discovery, regulators may consider a "likely validity" date since the blockchain can attest \textit{when} the original QUOTE was signed. 

Recall as well that the computation is auditable at any time: should the regulator question a result, they can simply trigger an audit of the private data, which could be as simple as running the OSC again with the auditor's oversight.  Consider as well that it is often vastly easier for a malicious data owner to forge their private data (which they can do today without AGAPECert) than it is to attempt cracking open a CPU in an attempt to probe for highly guarded embedded attestation keys.  Once such foul play is discovered during any audit, Intel EPID can simply revoke the key that the malicious data owner spent so much effort attempting to learn, providing severely diminishing returns to any such attacker.

Therefore the traditional side-channel security vulnerabilities with TEE computing have little ability to impact the security of AGAPECert in general.

%If the client is outsourcing the compute engine to run in a potentially untrusted environment, the OSC’s they use must implement techniques to protect against data leakage to malicious peers on the untrusted platform by moving all decrypted private data interaction inside the trusted enclave, as well as protecting against side channel attacks using techniques such as PATH ORAM~\cite{OleksenkoO.2020VPSe,Ahmad_Joe_Xiao_Zhang_Shin_Lee_2019,Stefanov2013}. Our future work can include a mix of Intel SGX and Path ORAM as a prominent tool to provide oblivious computation on a untrusted node (outsourced compute engine.) 

%-------------------------------------------------------------------------------
\subsection{Analyzing AGAPECert's Trust Levels}
\label{ssec:trust_levels_vulnerabilities}
%-------------------------------------------------------------------------------

\subsubsection{Owner Attested} 
The regulator trusts the data owner to correctly execute the OSC.  Therefore, the Compute Engine, Broker, Data Store are assumed to be trusted. In the case of an adversarial data owner, the regulator can still validate the private data and code execution via audits.

\subsubsection{Enclave Attested} 
Trust Level 2 requires the correct execution of the OSC/Algorithm. Therefore, Trust Level 2 relies on the attestation capabilities of the Intel SGX Architecture. Following our previous discussion on side-channel attacks, an adversarial data owner can steal secrets from an outdated Intel SGX-enabled node signing code and data as genuine compromising ultimately Trust Level 2. AGAPECert implementations must require up-to-date remote attestation schemes, including the latest SVN known to have reasonably uncompromisable attestation keys.

\subsubsection{Ordering Attested}
For Trust Level 3, the discussion about faithful code execution is analogous to Trust  Level 2.  Ordering of events provided by a shared ledger can expose the execution timestamp of an OSC and creation of a PAC breaking our premise of obliviousness and revealing useful information for an interested party. AGAPECert utilizes a Blockchain Gateway to submit \textit{anonymous} and \textit{asynchronous} transactions to a shared ledger or a mix of ledgers hiding the identity of the participants. In the case of a compromised shared ledger, no useful information is derived by an attacker solely from the global state. 

%-------------------------------------------------------------------------------
\subsection{Discussion}
\label{ssec:security_discussion}
%-------------------------------------------------------------------------------
There exist extremely sensitive datasets and data sources---electronic health records, stock market, finance, etc. ---in which even the leakage of a reduced set of bits can be catastrophic. Under adversarial environments, these use cases require stronger cryptosystems that offer semantic security~\cite{Gentry:2009:FHE:1834954,Paillier:1999:PCB:1756123.1756146} (homomorphic encryption). 
%(partially-homomorphic encryption, somewhat-homomorphic encryption, fully homomorphic encryption).
AGAPECert use cases and trust model limit the capabilities of an adversary.  For instance, a participant is limited by standards, regulations, and audits. Moreover, the possibility of legal action---when a deviation from the protocol is suspected---forces the participant to maintain a good reputation. 

%-------------------------------------------------------------------------------
%\subsection{Malicious Nodes in the Blockchain}
%\label{ssec:malivious_nodes_blockchain}
%-------------------------------------------------------------------------------
Powerful adversaries (stronger than HBC\footnote{"The honest-but-curious (HBC) adversary is a legitimate participant in a communication protocol who will not deviate from the defined protocol but will attempt to learn all possible information from legitimately received messages"~\cite{Paverd2014ModellingAA}.}) that can get access to the blockchain cannot derive any useful information from the stored cryptographic hashes and random universal unique identifiers. %For trust level 3, AGAPECert generates untraceable cryptographic hashes.
AGAPECert does not rely solely on Intel SGX attestation and sealing primitives; instead, AGAPECert contributes a set of trust levels providing adaptability features according to particular use cases. AGAPECert's future releases can allow homomorphic encryption exposing Homomorphic and Oblivious Smart Contracts (HOSC). Integrating SEAL~\cite{sealcrypto} in AGAPECert allows a richer set of devices as compute engines.

%-------------------------------------------------------------------------------
\section{Prototype Implementation}
\label{agapecert_sec:prototype_implementation}
%-------------------------------------------------------------------------------
The AGAPECert prototype implementation components are available as open source as shown in Table~\ref{tab:agapecert_components}, and documentation for how to install and run the entire flow can be found at https://github.com/agapecert/agapecert.

\begin{table*}[t]
\centering
\begin{tabular}{|p{2.6cm}|p{6.6cm}|p{6.9cm}|}
	\hline
	\textbf{Component} & \textbf{Repository} & \textbf{Objective}\\
	\hline
	Broker & https://github.com/agapecert/broker  & OSCs' Service Manager \\
	Validator  & https://github.com/agapecert/validator & PACs' verifier\\
	Compute Engine & https://github.com/agapecert/compute-engine  & Compute Engine\\
	Blockchain Gateway & https://github.com/agapecert/blockchain-gateway  & Anonymous and asynchronous shared ledger accesses\\
	
	\hline   
\end{tabular}
\caption{Summary of components' repositories for AGAPECert implementation.}
\label{tab:agapecert_components}
\end{table*}
%-------------------------------------------------------------------------------
\subsection{Blockchain Gateway}
\label{agapecert_ssec:fabric_implementation}
%-------------------------------------------------------------------------------

% revision 2021.09.19
For Trust Level 3, AGAPECert's prototype implementation interacts with a $pacContract$ deployed in a blockchain network through a custom Javascript-based Blockchain Gateway\footnote{https://github.com/agapecert/blockchain-gateway}.  The primary purpose of the Blockchain Gateway is to submit asynchronous (and optionally anonymous) %anonymous 
transactions to the ledger. The Blockchain Gateway is generalizable and can accommodate pluggable shared ledgers. This gateway allows the Broker, Validator, and OSC to interact with a shared ledger (or a mix of ledgers). AGAPECert's Blockchain-Gateway future releases can utilize concepts defined by Agrawal et al.~\cite{patent:20190164153} to enhance anonymity and privacy when interacting with a shared ledger.

%-------------------------------------------------------------------------------
\subsection{Trusted Compute Engine}
\label{ssec:compute_engine_implementation}
%-------------------------------------------------------------------------------
We implemented the OSCs utilizing C++ for the OSC exterior. We utilized OpenEnclave~\cite{OpenEnclave} to implement C bridge functions for the OSC interior. The compute engine exposes a rich API (C++ driver Listing~\ref{lst:c_driver}) to allow secure communication between the OSC, the graph-based API, and the Broker. The C++ driver implements secure WebSockets.

% Authenticated Blocked Graph API
\begin{lstlisting}[language=JavaScript, caption={Example OSC Exterior usage of Compute Engine C++ Trellis driver (https://github.com/agapecert/compute-engine)}, label={lst:c_driver}]
Trellis * objTrellis = new Trellis();
objTrellis->getPrivateData().wait();
m_private_records = objTrellis->m_private_records;
// AGAPECert C++ Driver - Compute Engine Methods
objTrellis->getUUID();
objTrellis->getToken();
objTrellis->getAuthorization();
objTrellis->getPrivateDataPath();
objTrellis->getPrivateData();
objTrellis->putPAC();
\end{lstlisting}

%-------------------------------------------------------------------------------
\section{Example Applications}
\label{agapecert_sec:example_applications}
%-------------------------------------------------------------------------------
There are an enormous number of potential applications for AGAPECert with its OSC+PAC model. This section shows a few non-trivial illustrative examples\footnote{A complete description of example applications can be found at https://github.com/agapecert/osc-definitions/wiki}. %\textbf{We need a smaller version of the applications in this section}.
%, but this list should by no means be considered exhaustive or even fully representative.
%There are an enormous number of potential applications for AGAPECert with its OSC+PAC model. Here we show a few illustrative examples, but this list should by no means be considered exhaustive or even fully representative.

% revision 2021.09.18

%------------------------------------------------------------------------------- this was included previous ICDCS
\subsection{Trust Level 1: Automated Sustainability Reporting}
\label{ssec:tl1_asr}
%-------------------------------------------------------------------------------
A Consumer-Facing Food Company, known as CFFC, wishes to create periodic sustainability reports for consumers to strengthen their brand.
However, many of the metrics they might report are dependent upon data sources outside their company: i.e., their suppliers or third-party contractors.  Consider just one metric: total energy consumed from renewable sources.  CFFC does not want to require its suppliers to send their energy bills to them each month.  Instead, CFFC creates an OSC that looks in Trellis first for an already-produced sustainability report, and if it finds one, it produces a PAC with just the “total energy consumed from renewable resources” number extracted from the existing report.  If it does not find one, it looks for any energy bills containing such numbers, adds them together for the year, and then produces a PAC with the resulting number. 

CFFC asks its suppliers to run this OSC in their AGAPECert instances and then share the resulting PAC with CFFC via an automated Trellis connection. CFFC is able to fully automate the creation of their own sustainability report without requiring the release of private, sensitive data from their suppliers. This relies on their supplier having such information already in a Trellis-conformant data store: in cases where it is not, the benefits of automating this process for their suppliers is incentive to achieve that goal in ways that are difficult to incentivize without such a tool.

The supplier has a trusted third party that handles running their OSC’s, so the third party starts up this OSC from CFFC, and an employee with the supplier uses the AGAPECert broker app to authorize, configure, and monitor the OSC as it creates the PAC’s automatically every month.  The employee creates an automated Trellis connection to CFFC so that whenever a new PAC is generated, and that PAC has been approved internally via a rule set or human approval, it automatically synchronizes the new PAC to CFFC. This represents a Trust Level 1 (Owner-Attested) PAC. The data owner is attesting that they ran the code faithfully, and the hash of the input data is stored in the PAC by the OSC, allowing auditability in the future as needed.

\subsection{Trust Level 2: Certified Fishing Catch Area}
\label{ssec:tl2_cca}
%-------------------------------------------------------------------------------

The global fishing industry would like to eliminate over-fishing by requiring fishing vessels to catch fish only in approved areas.  
%However, 
Fishermen consider their active catch areas to be proprietary to their business.
%: they don’t want their peers to rush in and steal their catch. 
The industry needs a practical zero-knowledge proof that can certify a particular fish was caught within legal boundaries without disclosing the actual catch locations. A sustainable fishing industry consortium creates an OSC which checks a given Trellis data store for a list of catch locations within a certain time period or within a certain group identifier like a lot number. The OSC pulls a set of approved geospatial catch boundaries from an industry list, intersects the catch locations with the boundaries, and produces a PAC with the time period or lot number, a “true/false” answer about whether all the catch locations fit within the boundaries, an identifier for the set of boundaries used, and a QUOTE from the trusted black box attesting that the OSC code was executed faithfully.

The industry would like to minimize cheating, and authorizes a set of trusted catch location recording devices that are maintained and periodically tested by approved auditors. The OSC can be augmented to verify that each catch location contains a signature from a trusted recording device manufacturer, and that the device is present in a recent active audit that is digitally signed by a trusted third-party auditor, all using the Trellis standard document integrity signature process.  This additional “true/false” answer is added to the PAC, verifying that the catch locations themselves were attested by trusted parties other than just the fisherman.
The fisherman uses the AGAPECert broker and their OSC-enabled Trellis platform to authorize and run the OSC, producing the certification back to their Trellis platform.  Before each batch of fish can be sold at the docks, the buyer must receive the PAC for that days’ catch.
%via some means convenient to them. 
This represents a Trust Level 2 (Enclave Attested) computation where the global fishing industry would like to know that the code was faithfully executed by the fisherman.
\subsection{Trust Level 3: Organic Mass Balance}
\label{ssec:tl3_omb}
%-------------------------------------------------------------------------------
One of the more difficult certification problems is characterized by a mass balance. In its simplest form, there is some mass of product that has been certified to be produced (either based upon the total inputs to the process, or based upon a human auditor’s assessment), and the industry would like to know that the seller of a product indeed can certifiably produce that amount of that product. If an organic farmer could receive a certification that they can or did produce 10 tons of organic apples, downstream buyers of those apples would like to know that the farmer has not re-used that 10-ton organic certification multiple times with multiple buyers, thereby selling potentially non-organic apples under an organic certificate. The farmer, on the other hand, does not want to put a list of their transactions into some shared database for buyers to check for validity since this could tip off buyers about how much inventory he has or how many sales he has made recently and to whom. The buyer needs a zero-knowledge-style proof that the certified product they are buying has not been sold under this certification to someone else. % revision 2021.08.29 added point

An industry consortium agrees on one or more blockchain platform(s) to act as a byzantine fault tolerant shared datastore.  The consortium produce an OSC which can look at a buyer's Trellis platform for a private ledger of sales. This ledger should be initiated with an additive transaction that is digitally signed by a trusted auditor (i.e., the auditor attests that the farmer has 10 tons of organic apples). The signature is verified by the OSC, and the balance is computed by subtracting any subsequent verifiable sales. The OSC finds a proposed new sales transaction in the Trellis data store, digitally signed by the buyer and the seller. The OSC verifies that the amount of the sale does not exceed the available balance, verifies the signatures, and then saves the transaction to the end of the private ledger. The OSC  produces a PAC indicating “success/failure” for the transaction, which can be automatically saved back to either the buyer’s Trellis data store, the seller’s, or both. If the buyer has their own private ledger, this PAC from the seller can serve as auditable, traceable proof to include in the buyer’s private ledger and add to their available inventory, all without disclosing any of the buyer's sales to any outside party.

As specified to this point, the protocol suffers from a double-spending attack where the seller simply maintains multiple private ledgers, providing different ones to the OSC depending on which customer they are selling to, thus enabling them to “spend” the same certified product more than once. To alleviate this problem, we introduce the concept of a one-time-use asymmetric key pair generated by the OSC and verified by a smart contract on the OSC’s chosen blockchain.  When the OSC is evaluating a proposed transaction, it accesses the seller’s Trellis datastore to retrieve the one-time-use private key from the previous transaction and an indicator of where to find the corresponding public key in the blockchain storage layer. The OSC checks the blockchain’s record for that key to see if has been marked as “used” yet or not. If it has not been used, then the OSC initiates a smart contract at the blockchain platform to mark it as “used,” which is only allowed by the smart contract when it verifies that the OSC can produce a signature with the private key corresponding to the public key in the chain. The OSC sees this successfully finish and then asks a different smart contract on the blockchain platform to store a new one-time-use public key and registers it in the blockchain along with a hash of the new ledger.  The OSC will fail the transaction if the ledger hash for the key it used does not match the hash of the current ledger it is updating.  The OSC stores the private key for this new one-time-use key pair for the next transaction.  The OSC also checks that each successful transaction in the private ledger has a corresponding “used” key recorded in the blockchain with matching ledger hashes.
%The blockchain layer in this case acts as an optimistic lock on the “used” status of the key: if a malicious seller tries to use the same key twice, the second time will fail in the OSC and the malicious seller will not be able to get any enclave-attested OSC to sign and record a valid transaction in his private ledger. 
%Note that since Hyperledger uses an execute-order-validate architecture, it is possible for a race condition to exist where two or more near-simultaneous requests to mark a one-time-use key as “used” could both return success to the OSC.  Therefore, a reasonable delay should be enforced between transactions to allow the network to settle, and each subsequent run of the OSC must continue to validate all prior one-time-use keys for success status. 
Note the importance of including the hash of the ledger in the chain with the public key.  We do not want to link transactions in the chain by allowing one key to point to the next key in the chain. However, 
if the two are truly unlinkable, then it is possible for a malicious seller to still double-spend by simply maintaining multiple ledgers with multiple one-time-use keys. Therefore, the OSC and smart contract must allow the link to be maintained in the private ledger, both refusing to create and store a new one-time-use keypair whose full ledger hash is already present with another one-time-use key in the chain.  For maximal trust, the blockchain nodes themselves should also be capable of performing remote attestation to validate a QUOTE from the OSC interior.
%, %thereby 
%allowing the smart contract itself to only allow actions from an OSC that can prove it is the correct codebase. The seller can maintain his own private ledger of sales without disclosing them to any industry body or any of his potential buyers. The buyers do not know how much of a given product he has left to sell, how many times he has already made sales, or who he is selling to. They only receive a PAC that attests “true” to the question of whether the seller has a private ledger which adheres to the protocol and has an appropriate available balance to complete the transaction. Also consider that since the one-time-use keys in the blockchain are not linked together, and they are not identifiable to any given instance of an OSC, then anyone with access to the blockchain cannot figure out which keys go on who’s private ledgers, they can only learn this information from the private ledger itself.
%comment
This is an example of a Trust Level 3 (Ordering Attested) computation.  %because the relative timing of transaction events is attested by the blockchain as an optimistic lock on the one-time-use keys.  
%This is an example of an enclave-attested computation because the chaincode should be able to verify that it is talking with a true enclave and that enclave is indeed running the OSC code expected by the blockchain’s smart contract, and a malicious seller is therefore unable to get the OSC code to generate an attested transaction on his private ledger.

%-------------------------------------------------------------------------------
\section{Evaluation}
\label{agapecert_sec:evaluation}
%-------------------------------------------------------------------------------
In order to present empirical evidence of AGAPECert effectiveness, we develop three experiments to benchmark critical components of AGAPECert.
%We evaluated AGAPECert through a series of experiments on commodity hardware and a trusted edge server.
%In this section, 
We start by evaluating the trusted compute engine performance with a mix of Trust Level 1 and Trust Level 2. Then, we discuss the results of generating $1000$ PACs for different input sizes (a complete workflow performance evaluation). Finally, we evaluate the Blockchain-Gateway with our $pacContract$ and Hyperledger Fabric performance. %\servio{Finally, we compare our technique against a C++ library for Verifiable Computation~\cite{libsnark}.} 
All experiments are run $1000$ times. All the code of experiments can be found at https://github.com/agapecert.

%-------------------------------------------------------------------------------
\subsection{Experiment setup}
\label{agapecert_ssec:experiment_setup}
%-------------------------------------------------------------------------------
To show AGAPECert usability and flexibility, we utilize a commodity HP Pavilion Laptop with an Intel Core i5-8250u CPU and 16GB of RAM running Ubuntu Linux $18.04$ LTS 64-bit Operating System (serves as a compute engine, graph data store, and Apache Spark server). We also use a trusted edge server, Dell R340, with an Intel Xeon E-2186G and 64GB of RAM running Ubuntu Server Linux $18.04$ LTS 64-bit Operating System (utilized for data center attestation primitives). The latter has support for Flexible Launch Control (FLC) and Data Center Attestation Primitives (DCAP.) Serving as the Blockchain-Gateway is a MacBook Pro (15-inch, 2017) with an Intel Core i7 2.8GHz and 16GB of RAM running macOS Catalina Version $10.15.4$ and IBM Blockchain Platform $1.0.31$ Visual Studio Code Extension.
%\begin{table}[t]
%\centering
%\begin{tabular}{|r|r|r|r|r|}
%	\hline
%	\textbf{n} & \multicolumn{3}{|c|}{HP Laptop} & \multicolumn{1}{|c|}{Dell R340} \\
%	&  JavaScript & NodeJS & Intel SGX & NodeJS \\
%	\hline
%	1M &     62.39  &    21.28 &   81.73 &   15.51 \\
%	2M &    124.56  &    42.67 &   97.78 &   31.12 \\
%    4M &    249.34  &    84.99 &  127.62 &   61.91 \\
%    8M &    499.01  &   170.25 &  189.97 &  123.87 \\
%    16M &   993.37  &   340.19 &  314.35 &  248.23 \\
%    32M &  1964.04  &   679.55 &  560.76 &  496.34 \\
%    64M &  3917.01  &  1361.15 & 1060.93 &  992.64 \\
%    128M & 7830.74  &  2723.15 & 2043.72 & 1987.39 \\
%    256M & 15636.90 &  5442.42 & 4029.08 & 3972.21 \\
%    512M & 31389.48 & 10892.74 & 7978.08 & 7950.09 \\
%	\hline   
%\end{tabular}
%\caption{Summary of Evaluation Results.}
%\label{tab:evaluation}
%\end{table}

%-------------------------------------------------------------------------------
\subsection{Trusted Compute Engine Performance}
\label{agapecert_sec:compute_evaluation}
%-------------------------------------------------------------------------------
We developed a computation-intensive algorithm---the Monte Carlo approximation---as an Oblivious Smart Contract. The Monte Carlo OSC was instantiated in AGAPECert's compute engines running NodeJS (TL1), in-browser JavaScript (TL1), and C using OpenEnclace API (TL2). Besides, we developed equivalent code in Python and deployed it in Apache Spark version $3.1.0$.

% revision 01 - 20211010 Figure -> Fig.
%-------------------------------------------------------------------------------
\subsubsection{AGAPECert's compute engine and Apache Spark}
\label{agapecert_sssec:tl12-spark}
\begin{figure}[ht]	
	\centering
	\includegraphics[width=\columnwidth]{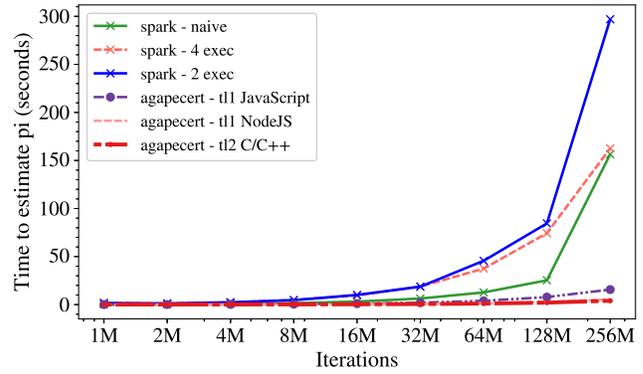}
	\caption{AGAPECert's compute engine (TL1, TL2) and Apache Spark (1,2,4 executors). We developed the Monte Carlo Approximation Algorithm for those compute engines.}
	\label{fig:agapecert_spark_compute_engine_performance}
\end{figure}

%-------------------------------------------------------------------------------
We compared AGAPECert's compute engine against Apache Spark version $3.1.0$. The goal of this experiment is to provide a context of comparison to AGAPECert; Apache Spark will scale and perform better at a massive scale. However, AGAPECert includes use cases in which preserving data ownership is critical.
Fig.~\ref{fig:agapecert_spark_compute_engine_performance} shows that AGAPECert's compute engine performs better than Apache Spark using this computationally-intensive Monte Carlo OSC. It is worth noting that Apache Spark's poor performance with two or four executors is due to the use of a synchronized random function in Python, which cannot scale to multiple cores. Nonetheless, a single executor (spark naive) provides a better comparison against AGAPECert's compute engine. Future AGAPECert releases can integrate Apache Spark~\cite{Zaharia2012} and Opaque~\cite{Zheng2017} for scalability guarantees.

%% ----------------------------------------------> Performance Plot

%-------------------------------------------------------------------------------
\subsubsection{Trust Level 1 and Trust Level 2 comparison}
\label{agapecert_sssec:tl12-comparison}
%-------------------------------------------------------------------------------

%% ----------------------------------------------> Performance Plot
\begin{figure}[ht]	
	\centering
	\includegraphics[scale=0.22]{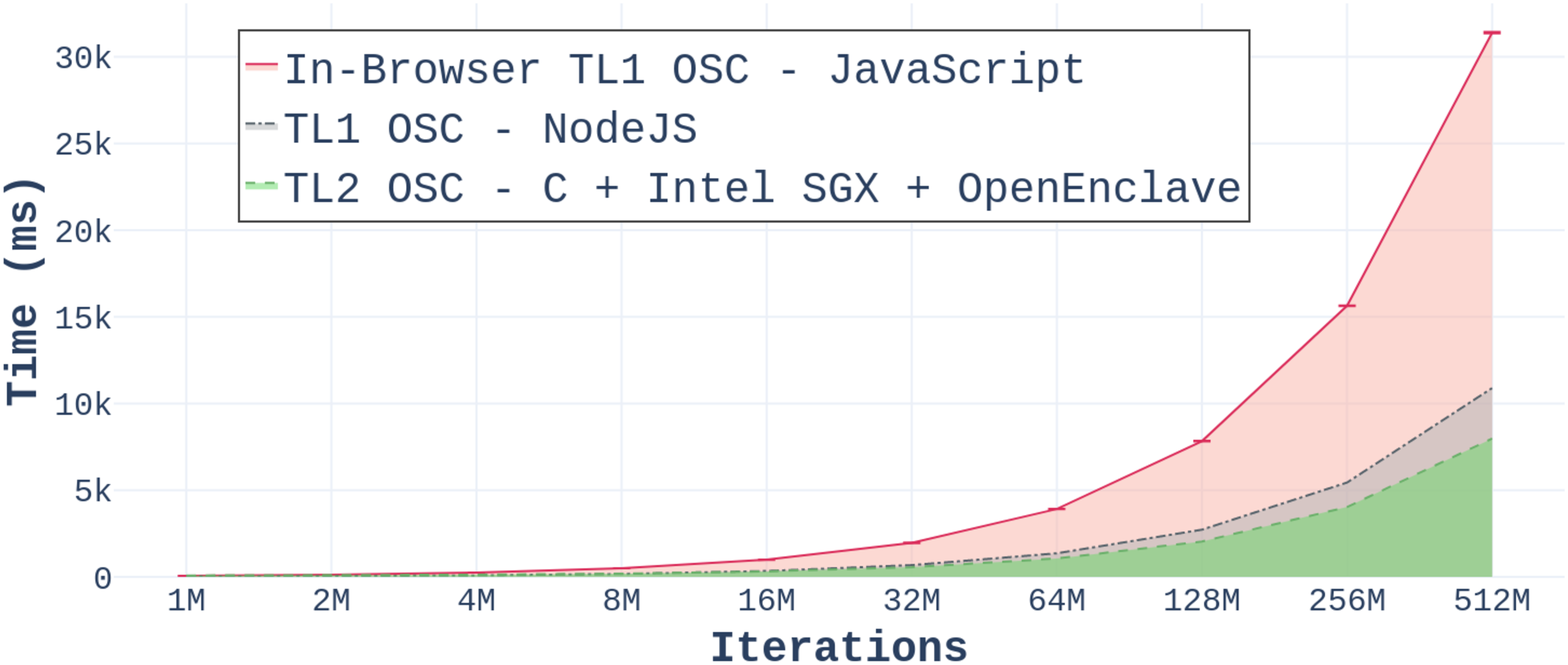}
	\caption{Compute Engine Performance. We developed the Monte Carlo Approximation Algorithm. The \textit{Monte Carlo OSC} is instantiated in AGAPECert for Trust Level 1 (TL1) for an in-browser compute engine, Trust Level 1 (TL1) NodeJS compute engine, and Trust Level 2 (TL2) with Intel SGX and OpenEnclave.}
	\label{fig:compute_engine_performance}
\end{figure}

% revision 01 - 20211010 Figure -> Fig.
In Fig.~\ref{fig:compute_engine_performance}, we observe that the AGAPECert (TL2) compute engine outperforms an in-browser JavaScript compute engine (TL1) with equivalent source code. Hence, AGAPECert (TL2) can offer similar performance to widely used development frameworks such as NodeJS and JavaScript. For some use cases, AGAPECert (TL2) can provide a better performance than such frameworks---even computing on top of encrypted private data. We compare the compute time exclusively. An extensive study of $ocalls$ and $ecalls$ performance is shown in~\cite{WeisseOfir2017Rlcw}.

%-------------------------------------------------------------------------------
\subsection{Private Automated Certifications Performance}
\label{sec:pac_evaluation}
%-------------------------------------------------------------------------------
This experiment shows the total time needed to generate a PAC using the K-means clustering algorithm deployed as an Oblivious Smart Contract (OSC). % review 2021.08.29 (OSC.) -> (OSC).
The K-means algorithm is setup with $n$ in increments of $2^i*1000$ where $i=0,1,2,3,4,5$; $k$ is set to buckets of $250$ items per cluster ($k=n/250$); $k$ centroids are determined randomly. The total time includes enclave creation, communication with the graph-based API, computation on private-data, PAC generation, and PAC storage in the graph data store. An asynchronous blockchain access stores the PAC in the blockchain (Fig.~\ref{fig:blockchaingateway_performance}). 

%% ----------------------------------------------> Performance Plot
\begin{figure}[ht]	
	\centering
	\includegraphics[scale=0.22]{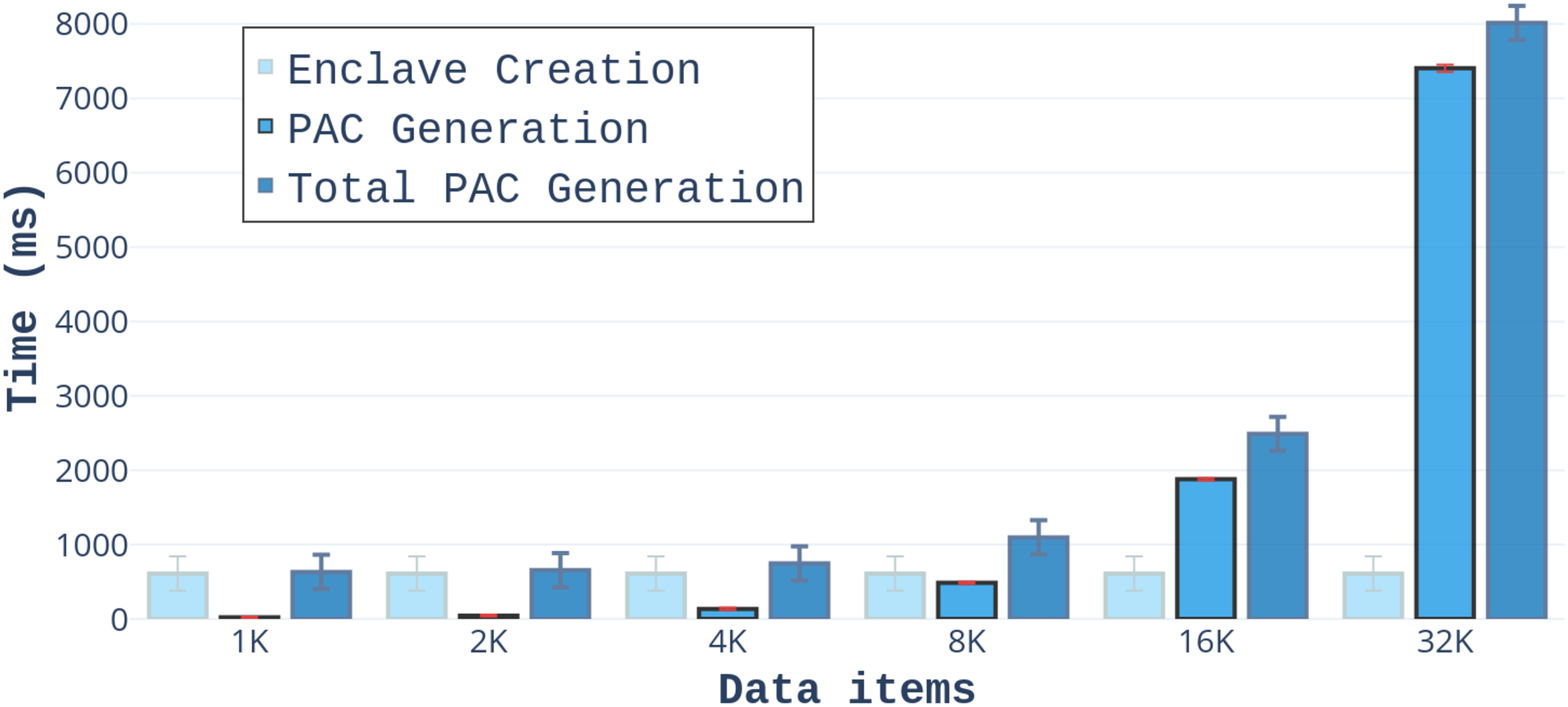}
	\caption{PACs' generation performance evaluation for Trust Level 3 (TL3). The K-means algorithm is instantiated as an Oblivious Smart Contract (\textit{K-means OSC or oblivious K-means if you will}).}
	\label{fig:pac_performance}
\end{figure}

The purpose of this \textit{K-means OSC} is two-fold: (1) it presents a widely used clustering algorithm for replicability; (2) solves our example applications (\S\ref{ssec:tl2_cca})---certified fishing catch area and similar use cases---with a straightforward modification. % review 2021.08.29 (\ref{ssec:tl2_cca}) -> (\S\ref{ssec:tl2_cca})
The regulator fixes the set of centroids, the algorithm to compute the distances, and the threshold that determines if the PAC has passed the evaluation/certification process. As mentioned before, this K-means OSC will be agreed upon by both the regulator and the data owner. The K-means OSC will contain all the semantics that allow the correct validation of data to generate an objective---according to the specification---PAC that can be audited in the future.

% revision 01 - 20211010 Figure -> Fig.
Fig.~\ref{fig:pac_performance} shows that the PAC generation is bounded by the input data retrieved from the graph data store. Additionally, the running-time in the blockchain (\S\ref{agapecert_sec:blockchain_gateway_evaluation}) and enclave creation is approximately constant ($610.65\pm230.10$ $ms$). However, an OSC is continuously running, waiting for signals from the broker (OSCs' service manager.) The overhead to initialize an enclave is suffered only once per computation cycle, or when a restart is required. 

%-------------------------------------------------------------------------------
\subsection{Blockchain-Gateway Performance}
\label{agapecert_sec:blockchain_gateway_evaluation}
\begin{figure}[ht]	
	\centering
	\includegraphics[scale=0.22]{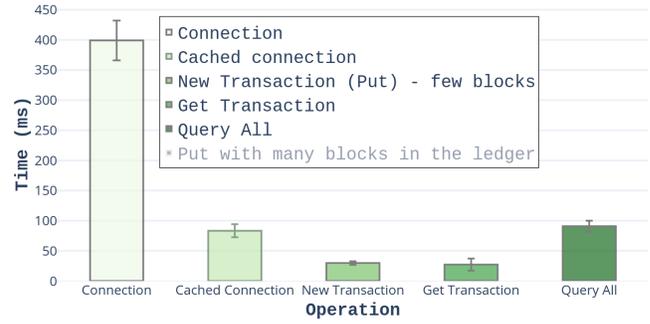}
	\caption{Blockchain-Gateway interacting with IBM Hyperledger Fabric and $pacContract$ performance.}
	\label{fig:blockchaingateway_performance}
\end{figure}

%-------------------------------------------------------------------------------
AGAPECert interacts with a Blockchain-Gateway for Trust Level 3~(\S\ref{agapecert_ssec:fabric_implementation}). We created a test suite to measure the Blockchain-Gateway performance using the Chai assertion library~\cite{Chai} and the Mocha test framework~\cite{Mocha} % review 2021.08.29 Figure -> Fig.
(Fig.~\ref{fig:blockchaingateway_performance}). When there exist multiple blocks in the shared ledger, the Blockchain-Gateway takes around $2180.88\pm38$ $ms$ to execute an asynchronous PAC creation in the ledger. The Blockchain-Gateway takes $27.37\pm9$ $ms$ to execute a PAC \textit{GET} query in the ledger (when there are more than 100 blocks in the shared ledger.) The validator will use the \textit{GET} function to the Blockchain-Gateway, the Broker or OSC will use the \textit{PUT} to the Blockchain-Gateway. The Broker, OSC, or validator will use the cached connection ($83.36\pm13$ $ms$).

%% ----------------------------------------------> Performance Plot

%-------------------------------------------------------------------------------
\section{Related Work}
\label{agapecert_sec:related_work}
%-------------------------------------------------------------------------------

This section describes prominent industry solutions that utilize Trusted Execution Environments to provide privacy-preserving computation. 

%-------------------------------------------------------------------------------
\subsection{Microsoft’s CCF framework}
\label{ssec:coco}
%-------------------------------------------------------------------------------

The Confidential Consortium Blockchain Framework (CCF, formerly CoCo)  strives to enable private and scalable blockchain networks~\cite{Coco}. The CCF is meant to be open-source and compatible with other blockchain protocols~\cite{Coco}. CCF augments the trust among peers/nodes executing smart contracts inside Intel SGX enclaves.  %The CCF integrates the global state and consensus algorithms with transactions and enclaves~\cite{br2018blockchain}.
Analogous to the CCF, we observe that traditional blockchains pose significant privacy issues; 
transactions, global state, and smart contract code are visible to anyone that enters the network. In contrast to the CCF, AGAPECert identifies that not all transactions and computation have to run in the chain for the problems addressed in this paper. 

%-------------------------------------------------------------------------------
\subsection{Secure Data Trading Ecosystem}
\label{ssec:sdte}
%-------------------------------------------------------------------------------

The Secure Data Trading Ecosystem (SDTE) aims to secure the data processing utilizing Ethereum and Intel SGX~\cite{8759960}. Analogous to AGAPECert, SDTE identifies the privacy implications when sharing complete datasets (sellers are data sources) with potential buyers or regulators. Instead, SDTE shares data analysis results processed on top of SGX-enabled nodes. Only trusted nodes execute data analysis contracts and the buyer can deploy any contract~\cite{8759960}. SDTE based their security guarantees mostly on remote attestation and sealing derived from the Intel SGX architecture. 
%Control flow inference attacks can detect and exploit vulnerabilities in SSL/TLS implementation in secure enclaves~\cite{XiaoYuan2017SDAS}. These attacks are particularly concerning for a computation running in the chain. 
CCF also utilizes Ethereum and Intel SGX to provide scalable and confidential blockchain networks~\cite{Coco}. Both solutions compute on-chain, whereas AGAPECert uses \textit{off-chain} business logic and certified pre-approved code---Oblivious Smart Contracts---that obtain only specific results from the private data.

%-------------------------------------------------------------------------------
\subsection{Secure Energy Trading Ecosystem}
\label{ssec:tdsc_secure_energy}
%-------------------------------------------------------------------------------
Aitzhan and Svetinovic (2018) identified the problems of centralized infrastructures~\cite{tdscenergyAitzhanNurzhanZhumabekuly2018SaPi}. They provided a peer-to-peer decentralized token-based system that allows a secure energy trading ecosystem. They rely on blockchain technologies to build their proof of concept.
Analogous to~\cite{tdscenergyAitzhanNurzhanZhumabekuly2018SaPi}, AGAPECert identifies the problems of relying on a centralized third party; instead, AGAPECert uses DCAP (Data Center Attestation Primitives) to provide a customized remote attestation for particular use cases. Moreover, AGAPECert offers various levels of trust, providing a flexible and generalizable framework.

%-------------------------------------------------------------------------------
\subsection{DeepChain}
\label{ssec:tdsc_deepchain}
%-------------------------------------------------------------------------------
Weng et al. introduced DeepChain, a distributed and collaborative training framework for deep learning~\cite{tdscdeepchainWengJiasi2019DAaP}. DeepChain relies on the decentralized nature of blockchain technologies to provide a reward-based mechanism that can force participants to behave honestly~\cite{tdscdeepchainWengJiasi2019DAaP}. DeepChain's simulator uses Corda as a shared ledger~\cite{tdscdeepchainWengJiasi2019DAaP}. Unlike DeepChain, AGAPECert does not rely entirely on blockchain technologies to provide auditability of transactions. AGAPECert's trust model is fundamentally different from DeepChain since AGAPECert defines a trusted compute engine. DeepChain utilizes a fixed set of smart contracts, i.e., a trading contract and a processing contract. AGAPECert is generalizable, allowing a myriad of blockchain technologies through its blockchain-gateway ---i.e., for trust level 3--- and the data owner can instantiate multiple algorithms as OSC in the framework.

%-------------------------------------------------------------------------------
\subsection{CAFE}
\label{ssec:tdsc_cafe}
%-------------------------------------------------------------------------------
CAFE is a cloud-based solution that utilizes hypervisor-level mechanisms to protect the deployment and execution of applications~\cite{tdscdxu8322174}. CAFE enables confidential execution in the cloud and is fundamentally different from Intel SGX~\cite{tdscdxu8322174}. In contrast, AGAPECert avoids sending data or code to the cloud; moreover, certified code---oblivious smart contracts---controls the algorithms that are allowed to run on top of private data, protecting data ownership.

%-------------------------------------------------------------------------------
\subsection{Google’s Asylo framework}
\label{ssec:asylo}
%-------------------------------------------------------------------------------
Asylo is an open-source framework that shields the integrity and confidentiality of data and applications through a confidential computing environment~\cite{Asylo}. %Asylo aims to facilitate the development and portability of applications. 
Asylo's most important goal is to make confidential computing easy, and therefore could be used in future development within AGAPECert.

%-------------------------------------------------------------------------------
\subsection{Teechain}
\label{ssec:teechain}
%-------------------------------------------------------------------------------

Teechain exploits TEEs to provide a layer-two payment network~\cite{lind2017teechain}. Teechain contributes asynchronous blockchain accesses. Teechain executes off-chain payments on top of Bitcoin utilizing a peer-to-peer network of TEEs~\cite{br2018blockchain,lind2017teechain}. Teechain is closely related to AGAPECert; however, it is closely focused on providing an off-chain payment network specifically and provides no private data access layer such as the Trellis framework used by AGAPECert.

%-------------------------------------------------------------------------------
\subsection{DelegaTEE}
\label{ssec:delegatee}
%-------------------------------------------------------------------------------
DelegaTEE aims to secure fine-grained delegation of rights and resources utilizing broker delegation on top of TEEs without revealing access credentials to third parties~\cite{SchneiderMoritz2019SBDT}. Since DelegaTEE allows fine-grained delegation of services or resources by an owner to a borrower, it can be used to enhance AGAPECert, enabling a richer interaction of services in the Certification model. DelegaTEE can allow the sharing of aggregated data (checkpoints) in the case of on-site audits to verify data and computation.

% review 2021.09.12
%-------------------------------------------------------------------------------
\subsection{Verifiable Computation}
\label{ssec:verifiable_computation}
	Cryptographic protocols such as \textbf{Verifiable Computation} (VC) facilitate outsourcing expensive computation kernels to stronger worker nodes or cloud workers so that the data owner or client can verify the computation's result faster than if the data owner performs the computation itself~\cite{10.1007/978-3-642-14623-7_25,10.1145/2856449,10.1145/2641562,10.1145/3338466.3358925,184499,7163030}.
	%From this definition, we extract that 

    %The trust model for VC %Verifiable Computation 
	%is envisioned to outsource the algorithm/function and data into cloud nodes, %or shared nodes,
	%which can potentially compromise data ownership. 
	The objective of VC ---offloading computation while maintaining  verifiable results--- is 
different from AGAPECert's goal of private automated certification between an owner and regulator.
AGAPECert includes a data owner that computes on its private data in its own node (or trusted service provider) to ensure data ownership.
	%The AGAPECert trust model is significantly different from the problem that VC tries to solve---outsourcing the computation to a possibly untrusted environment and providing proof of the correct function evaluation~\cite{10.1007/978-3-642-14623-7_25,10.1145/2856449}. On the contrary, the AGAPECert trust model includes a data owner who computes on its private data in its own node (or trusted service provider) to ensure data ownership. AGAPECert utilizes Trusted Execution Environments that allow the verification of the code that is being deployed in a particular enclave. This remote attestation process provides evidence that the code is running in an Intel SGX enabled platform, properly instantiated, and with a known security configuration. Therefore, an Intel SGX enabled platform exposes the means to verify the code and the data (input parameters), %providing non-repudiation characteristics, 
	% 2022.04.22 particularly for a computation happening in trusted compute nodes. AGAPECert's model and for particular use cases,  the code that runs inside enclaves can also be considered as private and only deployed to known/trusted nodes. Our technique abstains from sending data to a public or even a private network and applies restrictions to the code that runs inside enclaves. Although VC does not require trusted hardware to ensure security against malicious server behavior~\cite{10.1145/2641562,10.1145/3338466.3358925,184499}, it was conceived to outsource the computation and data, potentially compromising data ownership.
	AGAPECert abstains from sending data to a public or even a private network and applies restrictions to the code that runs inside enclaves. % 2022.04.22 which provides the non-repudiation property. 
In our model, the code that runs inside enclaves can be considered as private as it is deployed to known/trusted nodes. Although VC does not require trusted hardware to ensure security against malicious server behavior~\cite{10.1145/2641562,10.1145/3338466.3358925,184499,10.1145/2856449}, it was envisioned to outsource computation~\cite{10.1007/978-3-642-14623-7_25, 10.1145/2856449}, and it is meant to send the data and algorithms to a powerful outsourced/cloud node, potentially compromising data ownership. 
	%Therefore, VC exposes a mechanism to verify computation in a provable adversarial environment, while AGAPECert preserves data ownership allowing the data owner to execute trusted algorithms on its own trusted nodes and providing a cryptographically secure proof that a consumer of a PAC (Private Automated Certification) can verify later. 
	
	%Cryptographic-based 
	Due to this, VC would introduce drawbacks in the AGAPECert setting. In particular, cryptographic-based VC techniques such as Quadratic Arithmetic Programs~\cite{10.1145/2856449} and solutions that utilize Fully Homomorphic Encryption (FHE)~\cite{YU2019372,10.1145/2856449} introduce large computation overhead~\cite{Costan2016,https://doi.org/10.48550/arxiv.2106.14253} for trusted data owners that compute on their platforms. %(\S\ref{agapecert_sec:verifiable_Computation_evaluation}). 
	%Therefore, VC adds more computing overhead than what we need in our setting. 
	To verify this, we conducted a small experiment comparing the implementation of AGAPECert to VC for a common matrix multiplication task (code available at https://github.com/agapecert).
	%; we find that AGAPECert outperforms a popular VC library (libsnark) by $113$x.
	For this experiment, we prepared two identical DCsv2-series Microsoft Azure Virtual Machines (VMs), and built the VC use case using the C++ \textit{libsnark} library~\cite{libsnark}. The algorithm we compare computes the inner product of two vectors of size $n=100$, using Rank-1 Constraint Systems (R1CS) as in arithmetic circuit satisfiability~\cite{10.1145/2856449}. We find that $libsnark$ VC takes $1.049\times10^{-1} \pm 2.1\times10^{-3}$ seconds to compute the inner product, while AGAPECert only requires $9.287\times10^{-4} \pm 9.959\times10^{-5}$ seconds. 
	%(\S\ref{agapecert_sec:verifiable_Computation_evaluation}).
	%The implementation and integration with VC are out of the scope of this work and included as future work.
	The performance benefits of using Intel SGX compared to VC cryptographic-based schemes has also been noted in recent work~\cite{https://doi.org/10.48550/arxiv.2106.14253}.

% review 2021.09.12
%-------------------------------------------------------------------------------
\subsection{Commitment schemes}
\label{ssec:commitment_schemes}
%-------------------------------------------------------------------------------

	Commitment schemes enable committing to a chosen value/statement (time {$t_i$}) while hiding the content from others~\cite{BaKaPa11}.  The commitment schemes can reveal the committed value later (time $t_k$, where $k>i$). Commitment schemes include the commit phase in which a value is chosen and committed and the reveal phase during which the sender reveals the value—the receiver verifies its authenticity. The commit phase can consist of a single message—called commitment. To preserve the hiding property, the value chosen cannot be known by the receiver. To safeguard the binding property, the sender can only compute the message chosen during the commit phase—this phase needs a single message from the sender to the receiver and a verification check performed by the receiver. Commitment schemes can be integrated with AGAPECert to provide proof that can be verified later. AGAPECert utilizes group signatures schemes to verify that a particular algorithm was executed inside an enclave—providing a binding property to a specific set of valid processors~\cite{IntelEPID2017,intelsgxdcap}. The integration of Commitments schemes in AGAPECert will act as a Zero-Knowledge Proof to expose Computational Verifiable PACs.
	
	An existing framework for the integration of VC with commitments is described in~\cite{7163030}, where Costello et al. introduce a Commit-and-Prove scheme into their Geppetto solution for ``Versatile Verifiable Computation". The authors identified the need for a protocol that ensures (i) the cryptographic material respects the semantics of the original program that they started with and (ii) that there are no mistakes during the compilation process~\cite{gepettoVideo}. In many real use cases requiring the privacy provided by TEEs, or other solutions such as FHE and Multi-Party Computation (MPC), participants are reluctant to provide even encrypted versions of sensitive data to a public cloud, as would be the case with VC and commitments. AGAPECert's trust model is unique from the current literature in this respect, utilizing TEEs in a trusted environment while providing automated code verification and private automated certifications (PACs).
	%The implementation and integration with Commitment schemes are out of the scope of this work and included as future work.
	
% review 2021.09.12
%-------------------------------------------------------------------------------
\subsection{Solutions based on Secret Sharing}
\label{ssec:solutions_based_secret_sharing}
%-------------------------------------------------------------------------------

    Many protocols based on secret sharing schemes~\cite{Shamir1979} can be used to solve distributed computing securely on data, this method significantly differs from the trust model and the problem that AGAPECert wants to tackle. For instance, Secure Multi-Party Computation allows a set of parties to perform secure distributed computing~\cite{10.1145/3387108}. MPC utilizes Secret Sharing as a building block.  A secret sharing scheme solves the problem of sharing a secret $S$ amongst a set $N$ of $n$ nodes/parties; the $n$ parties have to share pieces of the data, and any subset (threshold) of $t+1$ can reconstruct the secret $S$; however, no subset less than $t$ shares can reconstruct or learn anything about the secret $S$. On the contrary, the AGAPECert Trust model includes a trusted data owner who computes on its private data to ensure data ownership. AGAPECert abstains from sending data to a public or even a private network and applies restrictions to the code that runs inside enclaves. Due to an increasing set of tools for Secure Computation utilizing Trusted Hardware, implementing algorithms inside black boxes is straightforward and increases their deployment and utilization. On the other hand, Multi-Party Computation requires high expertise for the correct deployment of applications~\cite{10.1145/3387108}. 
    Moreover, since MPC stores data pieces in many locations, the participants will require communication devices, increasing deployment costs. Also, in MPC, malicious participants in the protocol are pre-assumed, which differs from AGAPECert's trust model in which the environment is regulated, and the data owner is considered trusted under threat of a legal challenge (if a deviation from the protocol is suspected).

%-------------------------------------------------------------------------------
\section{Conclusion and Future Work}
\label{agapecert_sec:conclusion}
%-------------------------------------------------------------------------------

This paper presented AGAPECert, an auditable, generalized, privacy-enabling certificate framework that protects the confidentiality of data, participants, and code. AGAPECert utilizes a unique mix of blockchain technologies, trusted execution environments, and a real-time graph-based API to define for the first time Oblivious Smart Contracts (OSCs) that generate auditable Private Automated Certifications (PACs).   AGAPECert offers pragmatic performance and is generalizable to many use cases and data types. AGAPECert has a significant impact providing an open source~\cite{AgapeCert2020} framework that can be adopted as a standard in any regulated environment to keep sensitive data private while enabling an automated workflow. 

% revision 2021.09.19
Due to AGAPECert's malleable architecture, our technique can be easily extended to provide additional features. For instance, AGAPECert's Blockchain-Gateway future releases can utilize concepts defined by Agrawal et al.~\cite{patent:20190164153} to enhance anonymity and privacy when interacting with a shared ledger. Similarly, AGAPECert's future releases can allow homomorphic encryption exposing Homomorphic and Oblivious Smart Contracts (HOSC). Integrating SEAL~\cite{sealcrypto} in AGAPECert will allow a richer set of devices as compute engines, i.e., sensors, IoT devices, mobile devices, to name but a few. AGAPECert's roadmap includes analyzing and implementing techniques such as VC and Commitment schemes that can act as Zero-Knowledge Proofs. Finally, we will also analyze the integration of an automated source code repository for Oblivious Smart Contracts that can preserve the privacy and ownership of the source code.

%-------------------------------------------------------------------------------
\section{Acknowledgement}
\label{sec:acknowledgement}
%-------------------------------------------------------------------------------
Sponsorship for this work was provided by Foundation for Food and Agriculture Research (FFAR) under award 534662.

\ifCLASSOPTIONcaptionsoff
  \newpage
\fi



%\bibliographystyle{IEEEtran}

\bibliographystyle{ieeetr}
\bibliography{./main.bib}{}


%\newpage
\begin{IEEEbiography}[{\includegraphics[width=1.1in,height=1.5in,clip,keepaspectratio]{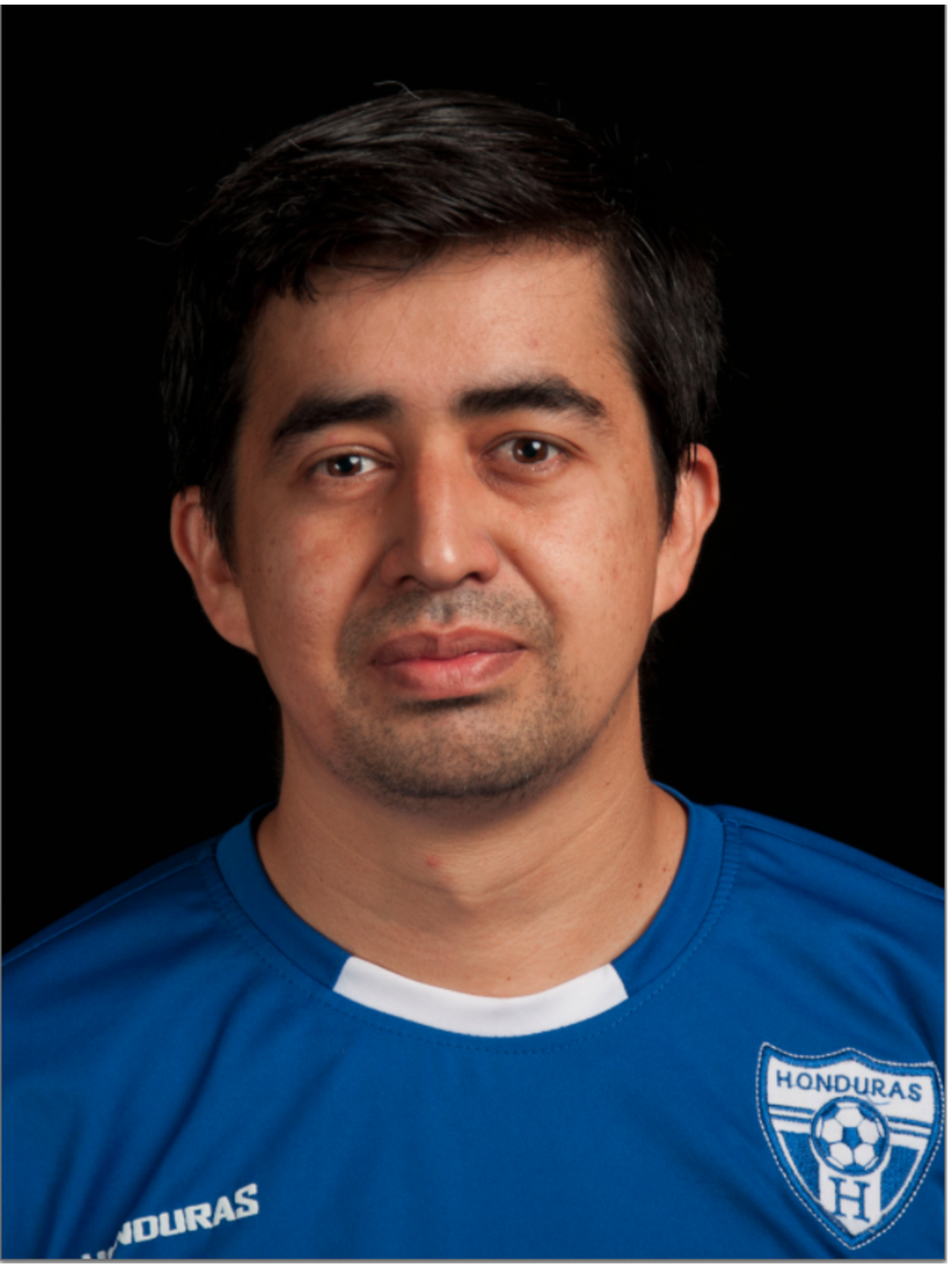}}]{Servio Palacios} is a Fulbright Scholar and a Ph.D. candidate with the Department of Computer Science at Purdue University. He holds an MSc in Computer Science from Purdue University, %a Master in Information Security from Foro Europeo de Navarra, 
and a B.S. in Computer Systems from UNITEC. Servio's research interests include graph theory, graph databases, distributed systems, edge computing, applied cryptography, operating systems, computer networks, distributed ledgers, blockchain, graph data science, confidential computing, and cybersecurity.
\end{IEEEbiography}
%\vskip -2\baselineskip plus -1fil
\begin{IEEEbiography}[{\includegraphics[width=1.1in,height=1.5in,clip,keepaspectratio]{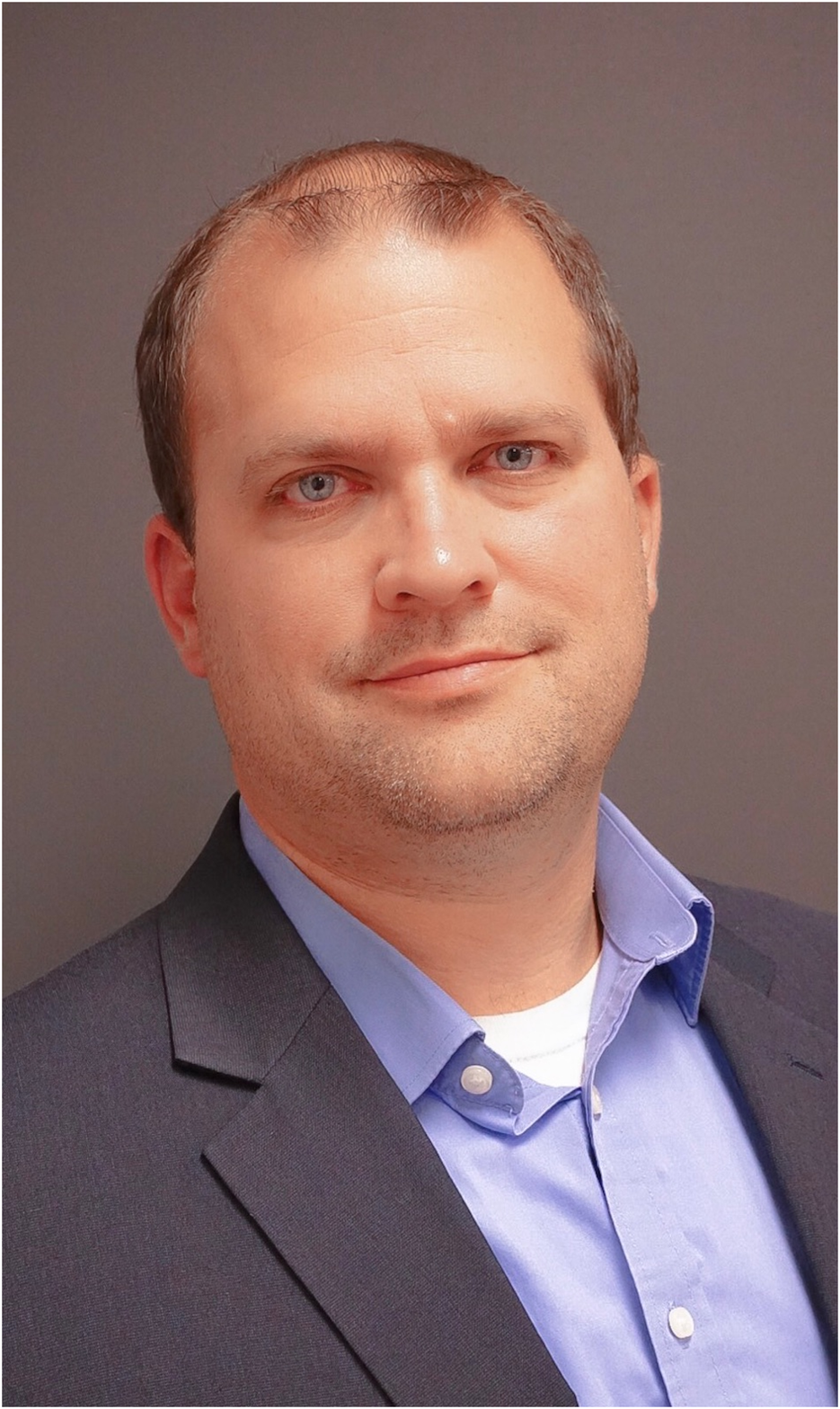}}]{Aaron Ault} is currently Senior Research Engineer with the Open Ag Tech and Systems Center (OATS) at Purdue University. He has contributed to projects across the computing and agricultural spectrum over the past 15 years, including signal processing, data science, embedded systems, software engineering, wireless sensor networks, scalable distributed cloud architectures, and mobile and web applications.  
\end{IEEEbiography}
%\vskip -2\baselineskip plus -1fil
%\newpage
\begin{IEEEbiography}[{\includegraphics[width=1.1in,height=1.5in,clip,keepaspectratio]{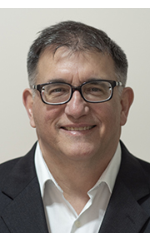}}]{James V. Krogmeier} received the BSEE degree from the University of Colorado at Boulder and the MS and Ph.D. degrees from the University of Illinois at Urbana-Champaign. He has industry experience in telecommunications, is a founding member of two software startup companies, and is the owner-operator of a Colorado wheat and corn farm. He is currently Professor and Associate Head of Electrical and Computer Engineering at Purdue University in West Lafayette, Indiana. Professor Krogmeier's research interests include the applications of statistical signal and image processing in agriculture, intelligent transportation systems, sensor networking, and wireless communications. 
\end{IEEEbiography}
%\vskip -2\baselineskip plus -1fil
\begin{IEEEbiography}[{\includegraphics[width=1.1in,height=1.5in,clip,keepaspectratio]{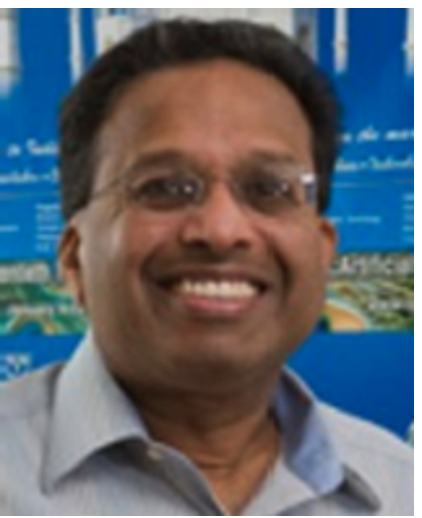}}]{Bharat Bhargava}
is a professor of the Department of Computer Science with a courtesy appointment in the School of Electrical and Computer Engineering at Purdue University. His research focuses on security and privacy issues in distributed systems. He is a Fellow of the Institute of Electrical and Electronics Engineers and of the Institute of Electronics and Telecommunication Engineers. In 1999, he received the IEEE Technical
Achievement Award for his decade long contributions to foundations of adaptability in communication and distributed systems. He serves on seven editorial boards of international journals. He also serves the IEEE Computer Society on Technical Achievement Award and Fellow committees.
\end{IEEEbiography}
%\vskip 2\baselineskip plus -1fil
\begin{IEEEbiography}[{\includegraphics[width=1.1in,height=1.5in,clip,keepaspectratio]{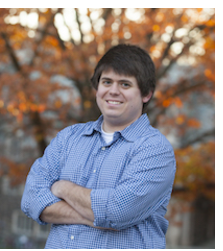}}]{Christopher G. Brinton (SM'20)}
is an Assistant Professor in the School of Electrical and Computer Engineering (ECE) at Purdue University. Previously he was the Associate Director of the EDGE Lab and a Lecturer of Electrical Engineering at Princeton University. Dr. Brinton's research interest is at the intersection of data science and networking, specifically in developing data-driven optimization methodologies for communication and social networks. He is a co-founder of Zoomi Inc., a big data startup company that has provided employee performance optimization for more than one million users, and a co-author of the book \textit{The Power of Networks: 6 Principles That Connect our Lives}.
\end{IEEEbiography}


%\vfill


\end{document}